\documentclass[12pt,preprint]{aastex}

\newcommand{\kms}{km s$^{-1}$}
\newcommand{\ha}{H$\alpha$}
\newcommand{\solar}{\ifmmode_{\sun}\else$_{\sun}$\fi}

\newcommand{\HII}{H$\,${\sc ii}}
\newcommand{\HI}{H$\,${\sc i}}

\newcommand{\coldens}{atoms cm$^{-2}$}

\begin{document}

\title{A comparison of young star properties with local galactic environment for LEGUS/LITTLE THINGS dwarf irregular galaxies}

\author{
Deidre A.\ Hunter\altaffilmark{1},
Angela Adamo\altaffilmark{2},
Bruce G.\ Elmegreen\altaffilmark{3},
Samavarti Gallardo\altaffilmark{1,4},
Janice C.\ Lee\altaffilmark{5},
David O.\ Cook\altaffilmark{6},
David Thilker\altaffilmark{7},
Bridget Kayitesi\altaffilmark{7,8},
Hwihyun Kim\altaffilmark{9},
Lauren Kahre\altaffilmark{10},
Leonardo Ubeda\altaffilmark{5},
Stacey N.\ Bright\altaffilmark{5},
Jenna E.\ Ryon\altaffilmark{5},
Daniela Calzetti\altaffilmark{11},
Monica Tosi\altaffilmark{12},
Kathryn Grasha\altaffilmark{11},
Matteo Messa\altaffilmark{2},
Michele Fumagalli\altaffilmark{13},
Daniel A.\ Dale\altaffilmark{14},
Elena Sabbi\altaffilmark{5},
Michele Cignoni\altaffilmark{15,16},
Linda J.\ Smith\altaffilmark{17},
Dimitrios M.\ Gouliermis\altaffilmark{18,19},
Eva K.\ Grebel\altaffilmark{20},
Alessandra Aloisi\altaffilmark{5},
Bradley C.\ Whitmore\altaffilmark{5},
Rupali Chandar\altaffilmark{21},
Kelsey E.\ Johnson\altaffilmark{22}
}

\altaffiltext{1}{Lowell Observatory, 1400 West Mars Hill Road, Flagstaff, Arizona 86001 USA}
\altaffiltext{2}{Department of Astronomy, The Oskar Klein Centre, Stockholm University, Stockholm, Sweden}
\altaffiltext{3}{IBM Research Division, T.\ J.\ Watson Research Center, Yorktown Heights, New York 10598 USA}
\altaffiltext{4}{Northern Arizona University, Flagstaff, Arizona 86011 USA}
\altaffiltext{5}{Space Telescope Science Institute, 3700 San Martin Drive, Baltimore, Maryland 21218 USA}
\altaffiltext{6}{Department of Astronomy and Astrophysics, California Institute of Technology, Pasadena, California 91125 USA}
\altaffiltext{7}{Department of Physics and Astronomy, The Johns Hopkins University, Baltimore, Maryland USA}
\altaffiltext{8}{Hamilton College, 198 College Hill Rd, Clinton, New York 13323 USA}
\altaffiltext{9}{Gemini Observatory, Casilla 603, La Serena, Chile}
\altaffiltext{10}{Department of Astronomy, New Mexico State University, Las Cruces, New Mexico USA}
\altaffiltext{11}{Department of Astronomy, University of Massachusetts -- Amherst, Amherst, Massachusets 01003 USA}
\altaffiltext{12}{INAF -- Osservatorio di Astrofisica e Scienza dello Spazio, Bologna, Italy}
\altaffiltext{13}{Institute for Computational Cosmology and Centre for Extragalactic Astronomy, Durham University, Durham, United Kingdom}
\altaffiltext{14}{Deptartment of Physics and Astronomy, University of Wyoming, Laramie, Wyoming USA}
\altaffiltext{15}{Department of Physics, University of Pisa, Largo B. Pontecorvo 3, 56127, Pisa, Italy }
\altaffiltext{16}{INFN, Largo B. Pontecorvo 3, 56127, Pisa, Italy}
\altaffiltext{17}{European Space Agency/Space Telescope Science Institute, Baltimore, Maryland 21218 USA}
\altaffiltext{18}{Zentrum f\"ur Astronomie der Universit\"at Heidelberg, Institut f\"ur Theoretische Astrophysik, Albert-Ueberle-Str.\,2, 69120 Heidelberg, Germany}
\altaffiltext{19}{Max Planck Institute for Astronomy,  K\"{o}nigstuhl\,17, 69117 Heidelberg, Germany}
\altaffiltext{20}{Astronomisches Rechen-Institut, Zentrum f\"ur Astronomie der Universit\"at Heidelberg, M\"onchhofstr.\ 12--14, 69120 Heidelberg, Germany}
\altaffiltext{21}{Department of Physics and Astronomy, The University of Toledo, Toledo, Ohio 43606 USA}
\altaffiltext{22}{Department of Astronomy, University of Virginia, Charlottesville, VA 22904 USA}

\begin{abstract}
We have explored the role environmental factors play in determining characteristics of young stellar objects in nearby
dwarf irregular and Blue Compact Dwarf galaxies.
Star clusters are characterized by concentrations, masses, and formation rates,
OB associations by mass and mass surface density,
O stars by their numbers and near-ultraviolet absolute magnitudes, and \HII\ regions by \ha\ surface brightnesses.
These characteristics are compared to surrounding galactic pressure, stellar mass density,
\HI\ surface density, and star formation rate surface density.
We find no trend of cluster characteristics with environmental properties,
implying that larger scale effects are more important in determining cluster characteristics
or that rapid dynamical evolution erases memory of the initial conditions.
On the other hand, the most massive OB associations are found at higher pressure and \HI\ surface density,
and there is a trend of higher \HII\ region \ha\ surface brightness with higher pressure, suggesting
that a higher concentration of massive stars and gas are found preferentially in regions of higher pressure.
At low pressures we find massive stars but not bound clusters and OB associations.
%suggesting that star formation may proceed at lower efficiency in these regions.
We do not find evidence for an increase of cluster formation efficiency as a function of star formation rate density.
However, there is an increase in the ratio of the number of clusters to number of O stars with pressure, perhaps
reflecting an increase in clustering properties with star formation rate.
\end{abstract}


\section{Introduction} \label{sec-intro}

Age and mass distribution functions of compact star clusters are similar among different types of galaxies \citep[for example,][]{fall12,whitmore17}.
Nevertheless, the products of star formation come with a wide range in numbers of stars and spatial concentration \citep[e.g.,][]{maiz01}.
There are massive, compact, bound clusters, such as the old globular clusters and young super star clusters
\citep[e.g.,\ R136 in the LMC and clusters in NGC 1569, NGC 1705, SBS 0335-052, He 2-10:][]{r136,n1569,reines08,he210}.
Yet, spatially large or massive does not always mean spatially concentrated \citep{kc88,mex95}, as shown by large loose associations of stars,
some of which occupy a large fraction of a dwarf galaxy
\citep[e.g.,\ Constellation III in the LMC, IC 10, I Zw 18, VII Zw 403:][]{con3,ic10,izw18,viizw403}.
At the small end in spatial size or mass, the star formation process also produces both small compact clusters and small associations
\citep[e.g.,][]{aversa11,legus-cl}.

The characteristics of natal clouds affect the products of the star formation process.
Simulations by \citet{dobbs17}, for example, show that massive, dense, long-lived clouds produce massive clusters and that smaller clusters form
in short-lived clouds.
Furthermore, clustered star formation occurs in parts of clouds with enhanced turbulence and density while
isolated star formation is found in parts of clouds with subsonic turbulence \citep{evans99}.
In addition, star formation proceeds faster in higher density gas, leading to a more narrow age distribution, while there are
longer star formation time scales and larger age spreads in low density regions \citep{parmentier14}.

There is also evidence that the star formation rate (SFR) depends on pressure \citep{blitz06} and that
large-scale galactic conditions affect the star formation products \citep[see for example][]{whitmore17}. We see this spectacularly in merging
galaxy systems that have produced large numbers of super star clusters \citep[e.g., the Antennae system,][]{whitmore95}.
\citet{lada10}, for example, suggests that triggering cloud formation through processes that increase the pressure, such as shocks,
could be significant in interacting galaxies and in early galaxy formation when globular clusters formed.
The old halo globular clusters presumably formed at very high pressures \citep{kruijssen15,bruce17}.
In fact, in a study of a giant molecular cloud (GMC) in the Antennae system, \citet{antennae15} find that the pressure
in the region in which this GMC is embedded is $10^4$ times higher than that in a typical galaxy.
\citet{swinbank10} find that a massive starburst galaxy at $z=2.3$ has star-forming regions with luminosity densities
comparable to cores of GMCs but 100 times larger
and $10^7$ times more luminous than what we see locally.
High pressure regions caused by collision of supershells within a galaxy and external ram pressure stripping
can also facilitate star formation \citep{bernard12,egorov17}.
Furthermore, there is evidence that the fraction of star formation resulting in bound clusters, $\Gamma$, is higher
in regions of high SFR density, especially starburst systems \citep{adamo-haro11,goddard10,adamo-mrk930},
although this is not universally agreed on \citep{chandar15}.

Consequences of the environmental conditions on the star formation products may extend to more local regions within galaxies as well.
For example, \citet{adamo-m83} found that $\Gamma$ decreases by factors of a few from the center of M83 to the outer disk and
varies from region to region within the galaxy with the local SFR density. In addition the initial cluster mass function (ICMF)
steepens in the outer disk as the upper cluster mass limit declines. Changes in $\Gamma$ and the ICMF with radius are
consistent with a decrease in gas pressure \citep{adamo-m83} or, similarly, gas density \citep{kruijssen12}.
Streaming of gas around bar potentials and piling up at the ends can also be a local factor in creating giant star forming
regions \citep{ee80,renaud15}.
In Blue Compact Dwarfs (BCDs) the migration of gas to the central regions \citep{caroline00}
increases the SFR densities and the local pressure,
enhancing the ability to form massive, concentrated star clusters \citep{HE04}.

\citet[]{ee97} \citep[see also][]{ashman01}
have suggested that the dominant factor in determining the kind of unit (bound super star cluster, open clusters, associations)
that is formed is pressure. A high pressure environment
facilitates the formation of massive, bound clusters, whether the high pressure is the result of high gas density or large-scale shocks such as in merging galaxies.
An equally massive cluster could be born at lower pressure but it would not be bound.
This is consistent with finding super star clusters in starbursts and merging galaxies and not finding them today in galaxies like the Milky Way where star formation is driven by internal processes.
\citet{escala08}, on the other hand, argue that the formation of massive clusters is determined by gravitational instabilities and
surface density of gas, and so massive clusters represent the largest scale in galaxies not stabilized by rotation.
However, young and old super star clusters and the somewhat less massive populous clusters are found even in some nearby dwarf irregular (dIrr) galaxies
although dIrrs in general have gas densities too low for gravitational instabilities to drive star formation
\citep{bigiel10,eh15}.
On the other hand, these are usually systems in which there is evidence that the formation of massive star clusters is an anomalous event;
the galaxies are too small to sample the cluster mass function to the extreme of super star clusters \citep[e.g.,][]{n1569}.
In addition, super star clusters are also usually found in a setting of unusually high star formation activity overall, suggesting
that some external perturbation has produced large-scale flows \citep{mex95,billett02}.
%But for less extreme star-forming events, the pressure in normal \HII\ regions in dIrrs are higher than their surroundings by a factor of $\sim10$,
%likely the result of the gravitational field that comes from local gas concentrations \citep{eh00}.

Given the evidence that the characteristics of the SF products relate to their local galactic environment,
we undertook a study of five dIrr galaxies for which we have catalogues of star clusters and O star candidates
as well as pressure, stellar mass, integrated \HI, and \ha\ maps of the galaxies.
These data allow us to compare the concentrations and masses of young stellar objects %star clusters, OB associations, and distributions of O stars
with respect to the surrounding pressure, \HI\ surface density, and stellar mass surface density.
We use the term ``cluster'' here as defined by the LEGUS project \citep{legus-cl}: compact and centrally-concentrated sources (class 1 or 2 objects)
which could be gravitationally bound systems, as well as objects
with asymmetric profiles and multiple peaks on top of diffuse underlying wings (class 3).
We also identify larger and looser OB associations in our sample of galaxies, defined by clumping of O stars.
Thus, here we are probing the realm of ``normal'' star formation products, that is, not super star clusters,
and examining all types of SF products, from small compact clusters to larger OB associations, to individual O stars.
And, we are examining the role local factors play in determining the characteristics of these objects.

In Section \ref{sec-data} we describe the galaxy sample, the data that we worked with, and the way in which we have
defined the environment within the galaxies.
In Section \ref{sec-cl} we present the observational results for the  star clusters, in Section \ref{sec-hii} the results for \HII\ regions,
and in Section \ref{sec-ob} the results for the O stars, including
their characteristics as a function of galactic characteristics and characteristics of regions at different
pressures.
We also define clusterings of O stars and discuss the properties of these associations in Section \ref{sec-obassoc}
%In Section \ref{sec-rat},
and we compare the numbers of clusters and of O stars by pressure.
%In Section \ref{sec-discuss} we discuss what the observational findings imply about SF in dIrr galaxies, and
In Section \ref{sec-summary} we summarize our findings.

\section{Data} \label{sec-data}

\subsection{\it Galaxy sample}

There are 5 dIrr galaxies in common between the {\it Hubble Space Telescope} ({\it HST}) Legacy Extragalactic UV Survey\footnote[22]{Based on observations obtained with the NASA/ESA Hubble
Space Telescope, at the Space Telescope Science Institute, which is operated by the Association of Universities for Research
in Astronomy, Inc., under NASA contract NAS 5-26555.}
\citep[LEGUS,][]{legus}
and LITTLE THINGS\footnote[23]{ Funded in part by the
National Science Foundation through grants AST-0707563, AST-0707426, AST-0707468, and
AST-0707835 to US-based LITTLE THINGS team members and with generous technical and logistical support from the
National Radio Astronomy Observatory.} \citep[Local Irregulars That Trace Luminosity
Extremes, The \HI\ Nearby Galaxy Survey,][]{lt}.
LEGUS is an {\it HST} Cycle 21 Treasury survey aimed at exploring star formation from scales of
individual stars to kpc-size structures with multi-band imaging of 50 galaxies within 16 Mpc.
The galaxies span the range of star-forming disk galaxies, including dIrrs.
The LITTLE THINGS survey is a multi-wavelength survey
from the far-ultraviolet to 21-cm \HI\ emission
of 37 dIrr galaxies and 4 BCDs.
LITTLE THINGS is aimed at understanding what drives star formation in tiny systems.
The LITTLE THINGS galaxies were chosen to be nearby ($\leq$10.3 Mpc), contain gas so they could form stars,
and cover a large range in dwarf galactic properties.
The galaxies in common between these two surveys - DDO 50, DDO 53, DDO 63, NGC  3738, and Haro 29 -
are the focus of this study. Some basic properties of the galaxies are given in Table \ref{tab-sample}.
Haro 29 is classified as a BCD and NGC 3738 has similar characteristics; both are more extreme in their star-forming
properties compared to the other three systems and NGC 3738 is more extreme than Haro 29.

Table \ref{tab-sample} also lists the distances adopted for this study and the references from which they came.
These distances are used in order to be consistent with the rest of the LEGUS studies \citep[see][]{legus}.
We note, however, that recent photometry of stars in the LEGUS galaxies have also yielded distance measurements from the apparent
brightness of the tip of the red giant branch \citep{sabbi18}. A significant difference exists between the new LEGUS
distances and the referenced distances for NGC 3738 (5.3$\pm$0.3 Mpc vs.\ 4.9 Mpc)
and Haro 29 (3.4$\pm$0.3 Mpc vs.\ 5.9 Mpc).
\citet{tully13}  also give a distance of 5.3 Mpc for NGC 3738, while for Haro 29 they give a distance of 5.7 Mpc, which is close
to the distance we adopt here.
Furthermore, \citet{dist-d53} gives a distance of 3.42 Mpc for DDO 50 compared to the 3.05 Mpc that we adopt from \citet{dist-d50}.
If these distances were used here instead of those given in Table \ref{tab-sample}, the masses and brightnesses
of objects in NGC 3738 would be 1.2$\times$ higher, in Haro 29 would be 0.3$\times$ lower, and in DDO 50 would be 2.0$\times$ higher.
However, the relative comparison of objects would stay the same.

\begin{deluxetable}{lcccccccc}
%\tabletypesize{\small}
\tabletypesize{\tiny}
%\rotate
%\tablenum{1}
\tablecolumns{9}
\tablewidth{460pt}
\tablecaption{The Galaxy Sample \label{tab-sample} }
\tablehead{
\colhead{} & \colhead{} & \colhead{D} & \colhead{} & \colhead{M$_V$}
& \colhead{$R_D$\tablenotemark{c}}
& \colhead{log SFR$_D^{H\alpha}$\tablenotemark{d}} & \colhead{log SFR$_D^{FUV}$\tablenotemark{d}} \\
\colhead{Galaxy}  &  \colhead{Other names\tablenotemark{a}}
& \colhead{(Mpc)} & \colhead{Ref\tablenotemark{b}} & \colhead{(mag)}
& \colhead{(kpc)}
& \colhead{(M\solar\ yr$^{-1}$ kpc$^{-2}$)} & \colhead{(M\solar\ yr$^{-1}$ kpc$^{-2}$)}
& \colhead{$E(B-V)_f$\tablenotemark{e}}
}
\startdata
DDO 50     & UGC 4305, Holmberg II &  3.05 & 1  & -16.4 & $0.99 \pm 0.05$ & $-1.67 \pm 0.01$ & $-1.55 \pm 0.01$ & 0.028  \\
DDO 53     &  UGC 4459                     &  3.66 & 2  & -13.9 & $0.73 \pm 0.06$ & $-2.42 \pm 0.01$ &$-2.41 \pm 0.01$ & 0.034  \\
DDO 63     &  UGC 5139, Holmberg I  &  3.98 & 2  & -14.8 & $0.69 \pm 0.01$ & $-2.32 \pm 0.01$ & $-1.95 \pm 0.00$ & 0.045  \\
NGC 3738 & UGC 6565                      &  4.90 & 3  & -17.1 & $0.78 \pm 0.01$ & $-1.66 \pm 0.01$ & $-1.53 \pm 0.01$ & 0.009 \\
Haro 29     & UGCA 281                      &  5.90 & 4  & -14.7 & $0.30 \pm 0.01$ & $-0.77 \pm 0.01$ & $-1.07 \pm 0.01$  & 0.013 \\
\enddata
\tablenotetext{a}{Selected alternate identifications obtained from NED.}
\tablenotetext{b}{Reference for the distance to the galaxy.}
\tablenotetext{c}{$R_D$ is the disk scale length measured from $V$-band images.
From \citet{HE06} revised to the distance adopted here.}
\tablenotetext{d}{SFR$_D^{H\alpha}$ is the star formation rate, measured from
H$\alpha$, normalized to the area $\pi R_D^2$, where $R_D$ is the disk scale length \citep{HE04}.
SFR$_D^{FUV}$ is the star formation rate determined from {\it GALEX} FUV fluxes
\citep[][with an update of the  {\it GALEX} FUV photometry to the GR4/GR5 pipeline reduction]{HEL10}.}
\tablenotetext{e}{Foreground Milky Way extinction from \citet{galext}.}
\tablerefs{
1 -- \citet{dist-d50};
2 -- \citet{dist-d53};
3 -- \citet{dist-n3738};
4 -- \citet{dist-haro29}.
}
\end{deluxetable}

\subsection{\it Star cluster catalogues}

\begin{deluxetable}{lccccc}
\tablewidth{490pt}
\tabletypesize{\small}
\tablecolumns{5}
\tablecaption{Numbers of  clusters, OB associations, and O star candidates \label{tab-numbers}}
\tablehead{
\colhead{}  &  \colhead{\# clusters} &  \colhead{\# clusters} &  \colhead{\# clusters} & \colhead{} & \colhead{} \\
\colhead{}  &  \colhead{(masses $>10^3$ M\solar)} &  \colhead{(masses $>10^3$ M\solar)} &  \colhead{(no mass cut) } & \colhead{} & \colhead{} \\
\colhead{Galaxy}  &  \colhead{(ages$\le$10 Myrs)} &  \colhead{(ages$\le$100 Myrs)} &  \colhead{(ages$\le$100 Myrs)} & \colhead{\# OB assoc} & \colhead{\# O stars}
}
\startdata
DDO 50     & 7 & 11    & 56 & 17 & 404 \\
DDO 53     & 1 &  1    & 7 & 11 & 101 \\
DDO 63     & 0 &  0    & 12 & 6  & 105 \\
NGC 3738 & 51 & 138 & 172  & 3 & 281 \\
Haro 29     & 8 & 9     & 13  & 7 & 61 \\
\enddata
\end{deluxetable}

The sample galaxies were observed
with {\it HST}'s WFC3 imager and filters F275W, F336W, and F438W and the ACS imager with filter F814W.
In addition DDO 50, DDO 53, and DDO 63 were observed with ACS and filter F555W,
and NGC 3738 and Haro 29 were observed with ACS and filter F606W.
The LEGUS team developed an exhaustive procedure for identifying and checking resolved compact stellar clusters on the images,
and details of the catalogue preparation are given by \citet{legus} and \citet{legus-cl}.
The first step is an automatic identification using the algorithm SExtractor \citep{sextractor}.
The second step involves the imposition of science-driven criteria aimed at reducing false detections and
a visual inspection of selected clusters. This step imposes an absolute magnitude limit and therefore
the compact cluster sample misses low mass objects, especially at older ages.
Cluster catalogues for the LEGUS dwarf galaxies are presented by
Cook et al.\ (2018a, in preparation).

\begin{figure}[t!]
\epsscale{0.95}
\vskip -0.25truein
%\plotone{fig1.eps}
%\includegraphics[angle=0,width=1.0\textwidth]{fig21left.eps}
\includegraphics[angle=0,width=1.0\textwidth]{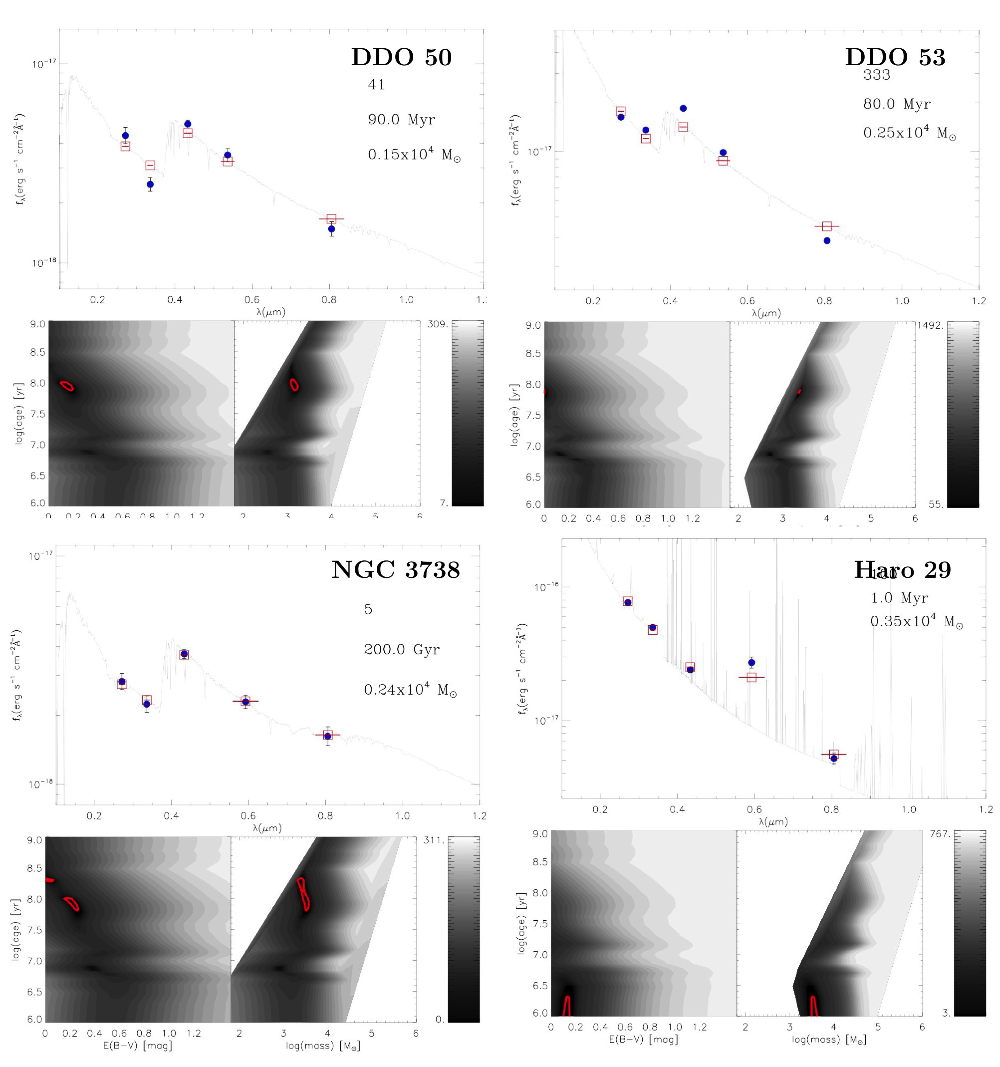}
\vskip -0.2truein
\caption{
Example SED fits for star clusters: cluster 41 in DDO 50, cluster 333 in DDO 53, cluster 5 in NGC 3738, and cluster 100 in Haro 29.
DDO 63 has no clusters. The process is described in detail in \citet{legus-cl}.
The top panel shows the flux vs.\ wavelength: observed fluxes as red squares and the best-fit spectrum as solid line and blue filled circles.
The age and mass are shown in the upper panel for each cluster.
The two panels below the SED panel shows the $\chi^2$ distribution in age, mass, and color excess E(B-V).
The scale is given to the right of the panels.
The red contours are the 68\% confidence level regions.}
\label{fig-clsed}
\end{figure}

Photometry of the clusters was performed in all available passbands.
The clusters were characterized by their Concentration Index
(CI), the integrated light within the central 1 pixel relative to that within a 3 pixel radius (pixel scale is 0.0396\arcsec).
Aperture corrections, as a function of filter, were made to the cluster photometry using two different methods:
1) taking an average of measurements of isolated clusters over an image
and 2) as a function of the CI of the cluster. Here we used both catalogues in the beginning, but found that it made little difference
to the results and subsequently adopted the CI-based aperture corrections.
Differences resulting from the two types of aperture corrections are discussed by %\citet{dwarfcl}.
Cook et al.\ (2018a, in preparation).
The photometry was corrected for Galactic extinction using \citet{galext} with the E(B-V) listed in Table \ref{tab-sample}.
Spectral energy distribution (SED)
fits were performed for those clusters with photometry in at least 4 filters in order to determine the age, mass, and reddening of the cluster within the host galaxy.
The fit for one cluster from each galaxy is shown as an example in Figure \ref{fig-clsed}.
Several internal reddening curves were used, and here we adopted the catalogues
in which the photometry was fit for internal extinction using the curve of \citet{starburstext}.
The SED fitting used two methods: 1) the {\it Yggdrasil} single stellar population models \citep{sed1}, and 2)
the stochastically sampled cluster evolutionary models of \citet{sed2}.
A \citet{kroupaimf} stellar initial mass function (IMF) from 0.1-120 M\solar\ was assumed.
%Our age and mass estimates do not take into account a stochastically sampled IMF of the stellar population within the clusters.
%However, a comparison between cluster physical properties derived with deterministic models and stochastically sampled IMF
%models shows very large deviations at cluster masses below 1000 M\solar\ \citep{sed2}.
The spread between stochastically based and deterministic derived cluster properties are within the age and mass uncertainties %\citep{dwarfcl}.
(Cook et al.\ 2018a, in preparation), but the differences are particularly noticeable for clusters with masses below 1000 M\solar\ \citep{sed2}.
More details on the production of the cluster catalogues are given by \citet{legus-cl}.

In using these catalogues, we eliminated clusters with masses less than 1000 M\solar\ in order to ensure completeness,
those observed in fewer than 4 filters (class 0), and those having a classification indicating that it is likely a foreground or background
source, single star, or artifact (class 4).
We included cluster classes 1, 2, and 3, where classes 1 and 2 are compact clusters and class 3 are more likely compact stelllar associations
\citep{legus-cl,grasha15,grasha17}.
We also imposed an age cut-off. We carried age cut-offs of 10 Myr, 50 Myr, and 100 Myr through the analysis, but differences were small,
so we present the results for 100 Myr below, except in the first comparison of cluster properties against environmental properties
we will also show the result for 10 Myr.
An age of 100 Myr minimizes losses due to dissolution of clusters \citep{lamers09,baumgardt13} \citep[but see][]{chandar17}.
without decimating the statistics of what are very small cluster samples.

\begin{figure}[t!]
\epsscale{0.7}
\vskip -0.2truein
%\plotone{fig2.eps}
\includegraphics[angle=0,width=1.0\textwidth]{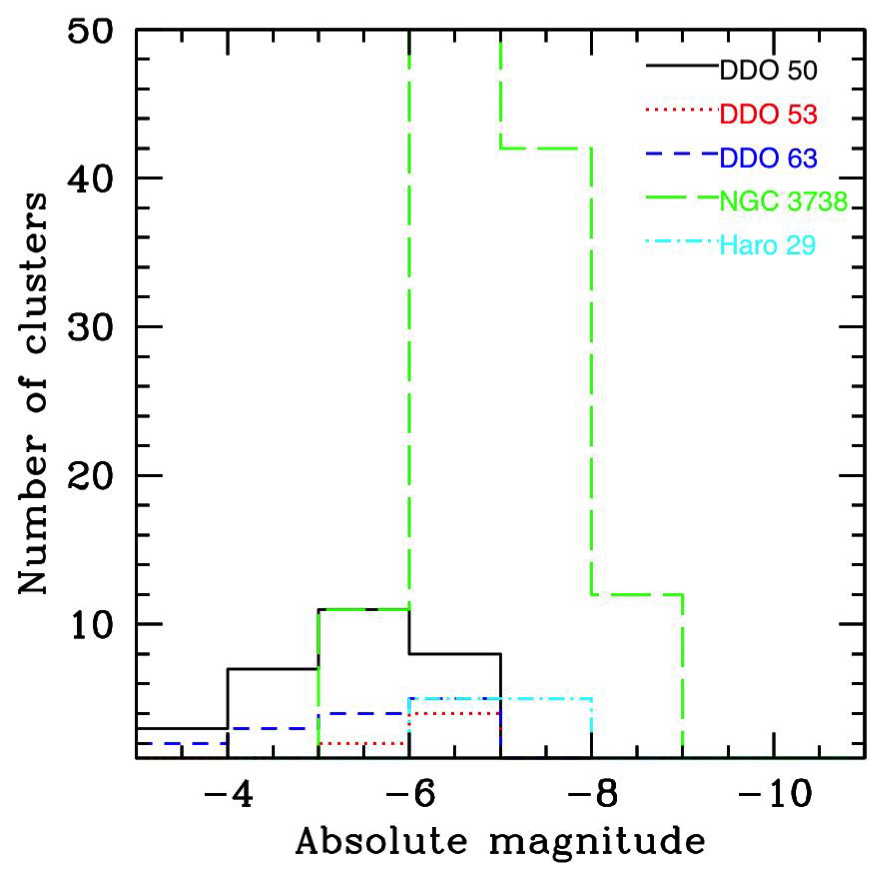}
\vskip -0.2truein
\caption{
Luminosity function for the clusters in our sample of galaxies. The absolute magnitude is in the $F555W$ filter for DDO 50,
DDO 53, and DDO 63, and in $F606W$ for NGC 3738 and Haro 29. NGC 3738 79 clusters in the first bin, but we have cropped
the y-axis at 50 in order to enable the other galaxy luminosity functions to be visible.}
\label{fig-lumfunc}
\end{figure}

\begin{figure}[t!]
\epsscale{0.7}
%\vskip -1.2truein
%\plotone{fig3.eps}
\includegraphics[angle=0,width=1.0\textwidth]{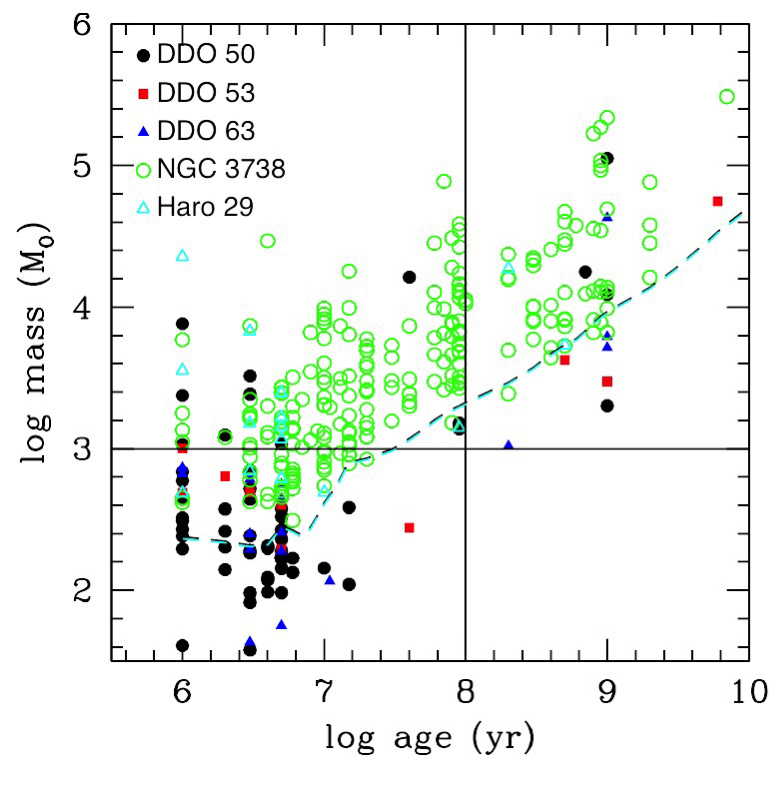}
%\vskip -4.7truein
\caption{
Mass versus age for the star clusters after cutting clusters from the catalogues on the basis of class (0 and 4) and
number of filters with observations ($<$4 filters). The vertical solid black line shows the age cut we apply throughout the analysis (age $>$100 Myr)
and the horizontal solid black line marks the cluster mass cut (mass $<$1000 M\solar).
The slanted dashed line marks the catalog limits for visual inspection of the clusters defined as $M_{F555W}=-6$ for DDO 50, DDO 53, and DDO 63
and $M_{F606W}=-6$ for NGC 3738 and Haro 29, a conservative estimate of completeness limit.
%We show the relationship for DDO 50 in black and for Haro 29 in cyan, as the galaxies with the smallest and largest distances in our sample.
%The relationship is nearly indistinguishable for the galaxies in our sample.
The concern for incompleteness is at the low mass ($<$2000 M\solar), older age ($>$35 Myr) corner of our selection box.}
\label{fig-massvsage}
\end{figure}

The clusters found automatically
were visually inspected in each galaxy down to a cluster absolute magnitude
$M_{\rm F555W}$ of $-6$, or for NGC 3738 and Haro 29 an $M_{\rm F606W}$ of $-6$.
We take the peak of the luminosity function of the clusters in a given galaxy as the 90\% cluster completeness limit.
In all but one LEGUS galaxy, the peak of the luminosity function is fainter than the limit for visual inspection of the clusters,
so a cluster absolute magnitude limit of $-6$ is a conservative indication of completeness.
In Figure \ref{fig-lumfunc} we show the luminosity functions for the clusters in our galaxies, and
in Figure \ref{fig-massvsage} we plot age versus mass for the clusters in our 5 galaxies before applying the mass and age cuts
that we use in the analysis.
The cluster absolute magnitude limit of $-6$ is translated here into age and mass, and shown as a slanting dashed line
from young, low mass clusters up to old, high mass clusters.
%Since the absolute magnitude limit is distance dependent, there is a different
%curve for each galaxy. However, the galaxies are at similar enough distances that the curves are nearly indistinguishable and we show
%only the curves for the closest galaxy, DDO 50, and the furthest, Haro 29.
We see that our $-6$ absolute magnitude limit shows potential incompleteness of clusters at the low mass, older age corner of our selection box
represented by our cut offs in mass and age.
Thus, in the analysis that follows, one should keep in mind that the clusters with masses $<$2000 M\solar\ and ages $>$35 Myr
may be somewhat more incomplete in this galaxy sample.

The numbers of  clusters are given in Table \ref{tab-numbers}, including the number of clusters before the cutoff for mass is applied
and the number of clusters with ages $\le$10 Myr.
There are relatively few clusters in DDO 50, DDO 53, and DDO 63 and even fewer with very young ages.
In fact, DDO 63 has no  clusters with masses $\ge$1000 M\solar, although it has 12  clusters
with smaller masses. DDO 63 does contain O stars.

\subsection{\it O star catalogues}

We also used LEGUS catalogues of candidate O stars in the dwarf galaxies %\citep{starcat}.
(Lee et al.\ 2018, in preparation).
These stars were selected to have magnitudes in NUV, $U$, $B$, and $V$ passbands,
have been flagged in the original stellar catalogues as having a point-source profile,
have an accurate F275W magnitude brighter than 25.5 (corresponding to a 3$\sigma$ detection),
and have a reddening-free parameter Q value greater than 1.6 with an uncertainty less than 0.075.
$Q$ is defined by $Q = (M_{F275W} -M_U) - K(M_U - M_B)$, where
$K$ is a constant that is computed using $E(NUV- U)/ E(U-B)$ with a Milky Way dust type ($R_V=3.1$).
Such a value of $Q$ selects for O stars, in particular stars with masses $\ge17$ M\solar.
Lower mass stars, having redder colors, have $Q$ values lower than this cut off.
However, O stars within \HII\ regions could be missed due to higher differential extinction.
Details on the LEGUS stellar photometry can be found in \citet{sabbi18}.

The magnitude cutoff of 25.5 mag in F275W is near the magnitude at which incompleteness starts to become severe
in these nearby galaxies.
However, in practice the faintest F275W mag in the catalogues are significantly brighter than this:
23.5, 22.6, 23.2, 22.7, and 22.5 in DDO 50, DDO 53, DDO 63, NGC 3738, and Haro 29, respectively.
Thus, the populations of stars we are working with are at least 2 mag brighter than the typical
limiting mag of 25.5 in F275W.
Hence, we are not likely to be significantly affected by incompleteness issues that might also bias the sample towards
more massive stars in more distant galaxies.
%(Galaxies vary from 3 Mpc to 6 Mpc distant).

The photometry is only corrected for foreground extinction according to \citet{galext}.
In our analysis where we use the F275W magnitudes of the stars, we apply an additional correction for internal extinction using an $E(B-V)=0.05$ mag
with the attenuation curve of \citet{starburstext}.
Actual extinctions are not known for our sample of galaxies, and so 0.05 mag is used as a
typical extinction for stars not buried in \HII\ regions. This value is half of the average $E(B-V)$ derived from
\HII\ region Balmer decrements for typical dwarfs \citep{HE06}, but
$E(B-V)$ could be larger in some dwarfs \citep{n4449}.
The numbers of candidate O stars are given in Table \ref{tab-numbers}.
%As for the compact clusters, since NGC 3738 and Haro 29 are more distant, we see below
%some evidence that fainter O stars are systematically missing and
%at the bright end the O star candidates could be blends.

\begin{figure}[t!]
\epsscale{0.75}
\vskip -1truein
%\vskip -0.25truein
%\plotone{fig4.eps}
\includegraphics[angle=0,width=1.0\textwidth]{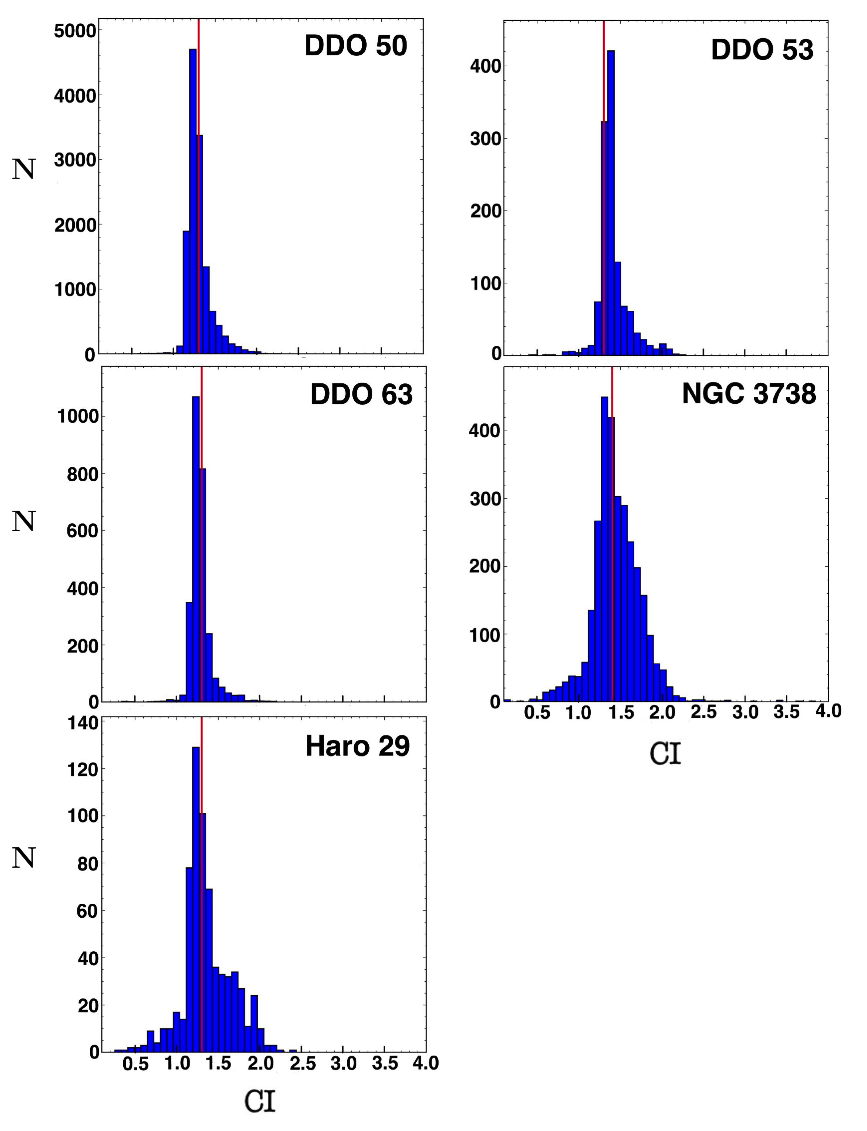}
%\vskip -1truein
\caption{
Number of stars and clusters as a function of CI for all sources identified in each galaxy.
The vertical red line denotes the CI marking the boundary between stars (lower CI) and clusters (higher CI) determined
from a training sample of unambiguous stars and clusters. See \citet{legus-cl} for more details
on the process of distinguishing stars from clusters used in the LEGUS sample.}
\label{fig-histCI}
\end{figure}

When we discuss the properties of star clusters and O stars below, one issue will be our ability to distinguish
compact clusters from stars. Our galaxy sample is relatively nearby and dIrr galaxies have lower stellar densities
than spirals, but we have approached this distinction carefully and systematically for the LEGUS sample as a whole.
The process is described in detail by \citet{legus-cl}, but the key step in distinguishing clusters from stars is
summarized here. The concentration index CI is determined for a  ``training sample'' of objects that are clearly stars and
those that are clearly clusters, and a histogram of the CI is plotted. There is a clear gap between the CI of stars
and the CI of compact clusters, and this is used to determine the CI cut for clusters and stars.
Figure \ref{fig-histCI} shows a plot of the CI for all sources extracted from our sample of galaxies, and the
red vertical line indicates the CI value used to distinguish compact clusters from stars in each galaxy.
One can see that even in the more distant galaxies in our sample (NGC 3738 and Haro 29) there is a fairly
clear drop to higher CI index. One can also see extended tails to higher CI values in each galaxy because
the clusters have a broader distribution of CI values than stars.

\subsection{\it OB associations}

Not all star-forming units are clusters or compact associations, and there are larger groups of O stars in all 5 galaxies.
Guided by the distribution of O stars as well as emission in the F275W images, we have, by eye,
outlined apparent OB associations. These are objects that appear as obvious density enhancements
in the number of O stars per area.
The size of an OB association is taken to be the radius of the circle that encompasses the O stars and F275W emission.
%that we are identifying as an OB association.

The northwestern part of NGC 3738 is problematic in regard to OB association identification; there are so many O stars and
clumps that it was difficult to decide whether it was a very large single association or a close grouping
of many individual clumps.
Whichever way it is described, it is an extraordinary region in terms of its size and density of O stars (and compact clusters), and
here we chose to emphasize that by considering it as a single region with a radius of 260 pc.
In addition most of the star formation in NGC 3738 is concentrated to this region, and we discuss the morphology of star formation
in NGC 3738 and a similar galaxy DDO 187 in \citet{2galpaper}.
If we had described that region instead as many smaller associations, what would we have found?
It is likely that most
of the smaller associations in the center of the region, while having smaller radii and stellar masses, would have similar
stellar mass densities and be associated with similar galactic environmental properties (see Figure \ref{fig-alln3738} below).
However, some of the smaller associations along the edges of the region in this scenario
would have lower mass densities and be described as having less extreme environmental properties as well.
It is unlikely that taking this alternate definition of the region in NGC 3738 would alter the results of the analysis presented here.

\begin{figure}[t!]
\epsscale{0.95}
\vskip -0.25truein
%\plotone{fig5.eps}
\includegraphics[angle=0,width=1.0\textwidth]{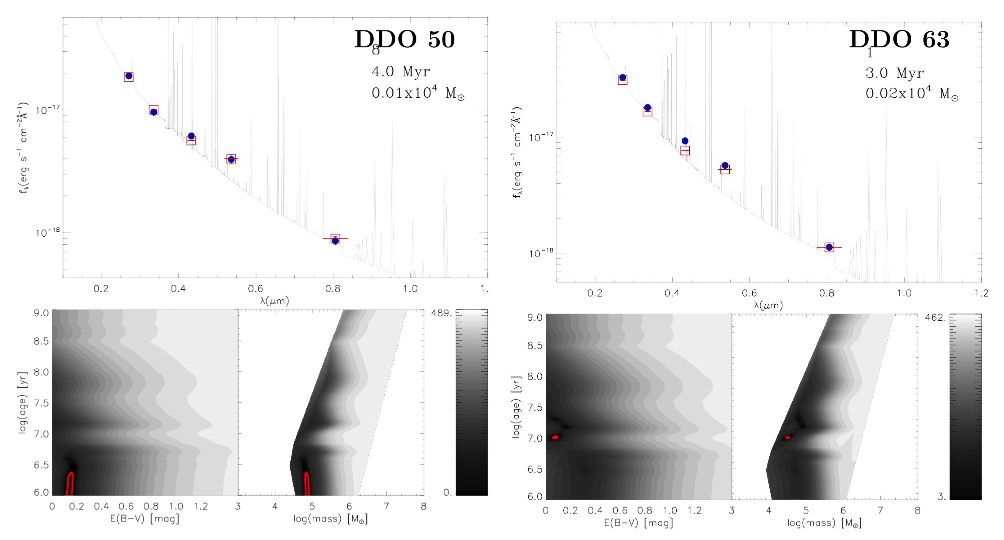}
\vskip -0.2truein
\caption{
Example SED fits for OB associations: \# 8 in DDO 50 and \#1 in DDO 63.
The top panel shows the flux vs.\ wavelength: observed fluxes as red squares and the best-fit spectrum as solid line and blue filled circles.
The age and mass are shown in the upper panel for each cluster.
The two panels below the SED panel shows the $\chi^2$ distribution in age, mass, and color excess E(B-V).
The scale is given to the right of the panels.
The red contours are the 68\% confidence level regions.}
\label{fig-obassocsed}
\end{figure}

We measured photometry in the circles encompassing the OB associations, chosen by eye to
include the O stars and F275W emission, on the {\it HST} images. % listed in Table \ref{tab-hst}.
We subtracted any  clusters within the boundary of the OB association and
subtracted background measured in an annulus around the association.
The photometry is on the Vega system for an infinite aperture size, and it was corrected
for foreground extinction E($B-V$)$_f$ according to \citet{galext}.
We performed SED fitting to the photometry, using the $Yggdrasil$ single stellar population models as for the  star clusters,
to determine the mass, age, and E($B-V$)$_i$ internal to the host galaxy.
Two examples of the SED fitting are shown in Figure \ref{fig-obassocsed}.
The OB associations are identified in Table \ref{tab-obassoc}, and the mass, the mass divided by the area of the encompassing circle,
age, and E($B-V$)$_i$ are given there.
For absolute F275W magnitudes, we applied the additional extinction correction
for internal reddening, as listed in Table \ref{tab-obassoc}, using an A$_{F275W}$/E($B-V$) of 7.43 \citep{starburstext}.
The goodness of the fit \citep[probability that the chi-square exceeds a particular value $\chi^2$ by chance,][]{recipes}
is also given in Table \ref{tab-obassoc}, where a
value of 1 denotes a good fit and a value near 0 means the fit is not well constrained.
Uncertainties in age and mass come from the maximum values and minimum values allowed by the fits.
The ages of the OB associations range from 1-50 Myr and radii of the encompassing circles from 20 pc to 300 pc.

\begin{deluxetable}{lccccrccccc}
\tablewidth{485pt}
\tabletypesize{\tiny}
%\rotate
\tablecolumns{11}
\tablecaption{OB associations \label{tab-obassoc}}
\tablehead{
\colhead{}  &  \colhead{} & \colhead{R.A. (2000)} & \colhead{Decl. (2000)} & \colhead{R} &
\colhead{} & \colhead{log Mass} & \colhead{log Age} &
\colhead{log Mass/Area} & \colhead{Goodness} & \colhead{} \\
\colhead{Galaxy}  &  \colhead{ID} & \colhead{(hh mm ss.s)} & \colhead{(dd mm ss)} & \colhead{(pc)} &
\colhead{$M_{F275W,0}$} & \colhead{(M\solar)} & \colhead{(yr)} &
\colhead{(M\solar pc$^{-2}$)} & \colhead{of fit} & \colhead{E(B-V)}
}
\startdata
DDO 50   &   1 &  8 19 17.4 & +70 43 42 &  43.4 & -11.65$\pm$0.002 & 3.90$^{-.07}_{-0.28}$ & 6.48$^{+0.12}_{-0.48}$ &  0.128 & 0.8 & 0.22$^{+0.04}_{-0.07}$ \\
         &   2 &  8 19 29.1 & +70 43  3 &  56.8 & -12.19$\pm$0.001 & 4.25$^{-.08}_{-0.28}$ & 6.30$^{+0.30}_{-0.30}$ &  0.243 & 0.9 & 0.11$^{+0.03}_{-0.05}$ \\
         &   3 &  8 19 29.1 & +70 42 51 &  53.9 & -11.76$\pm$0.001 & 3.97$^{-.10}_{-0.30}$ & 6.48$^{+0.12}_{-0.18}$ &  0.009 & 0.9 & 0.09$^{+0.03}_{-0.08}$ \\
         &   4 &  8 19 30.2 & +70 42 41 &  65.1 & -11.54$\pm$0.001 & 4.05$^{-.10}_{-0.28}$ & 6.00$^{+0.60}_{-0.00}$ & -0.074 & 0.8 & 0.07$^{+0.04}_{-0.04}$ \\
         &   5 &  8 19 26.5 & +70 42 48 &  60.4 & -11.42$\pm$0.002 & 3.82$^{-.08}_{-0.26}$ & 6.48$^{+0.12}_{-0.18}$ & -0.239 & 0.9 & 0.14$^{+0.04}_{-0.06}$ \\
         &   6 &  8 19 27.9 & +70 42 19 & 105.5 & -12.62$\pm$0.001 & 4.38$^{-.03}_{-0.22}$ & 6.60$^{+0.00}_{-0.60}$ & -0.164 & 0.9 & 0.04$^{+0.07}_{-0.03}$ \\
         &   7 &  8 19 29.3 & +70 42 29 &  45.7 & -11.08$\pm$0.004 & 4.06$^{+0.00}_{-0.23}$ & 7.00$^{+0.00}_{-0.00}$ &  0.243 & 0.0 & 0.23$^{+0.06}_{-0.05}$ \\
         &   8 &  8 19 12.2 & +70 43  8 &  94.9 & -13.58$\pm$0.001 & 4.86$^{-.08}_{-0.25}$ & 6.00$^{+0.30}_{-0.00}$ &  0.408 & 1.0 & 0.11$^{+0.05}_{-0.03}$ \\
         &   9 &  8 19 27.4 & +70 41 58 &  59.8 & -12.34$\pm$0.001 & 4.34$^{-.10}_{-0.25}$ & 6.00$^{+0.30}_{-0.00}$ &  0.290 & 0.9 & 0.14$^{+0.05}_{-0.04}$ \\
         &  10 &  8 19 25.9 & +70 41 53 &  59.2 & -10.89$\pm$0.002 & 3.73$^{-.08}_{-0.18}$ & 6.70$^{+0.00}_{-0.00}$ & -0.312 & 0.8 & 0.02$^{+0.03}_{-0.02}$ \\
         &  11 &  8 19 23.5 & +70 41 53 &  93.8 & -11.25$\pm$0.001 & 4.39$^{-.03}_{-0.19}$ & 7.18$^{+0.00}_{-0.03}$ & -0.051 & 0.4 & 0.03$^{+0.04}_{-0.03}$ \\
         &  12 &  8 19 23.0 & +70 42  3 &  64.5 &  -9.66$\pm$0.003 & 3.41$^{+0.36}_{-0.13}$ & 6.85$^{+0.27}_{-0.24}$ & -0.706 & 0.5 & 0.00$^{+0.23}_{-0.00}$ \\
         &  13 &  8 19 23.1 & +70 42 58 &  90.3 & -12.02$\pm$0.002 & 4.02$^{+0.18}_{-0.25}$ & 6.48$^{+0.70}_{-0.00}$ & -0.388 & 0.1 & 0.16$^{+0.04}_{-0.16}$ \\
         &  14 &  8 19 30.5 & +70 42 56 &  55.1 & -10.19$\pm$0.002 & 3.95$^{+0.02}_{-0.35}$ & 7.15$^{+0.03}_{-0.45}$ & -0.029 & 0.9 & 0.02$^{+0.16}_{-0.02}$ \\
         &  15 &  8 19 10.1 & +70 43 17 &  82.6 & -11.86$\pm$0.001 & 4.48$^{+0.39}_{-0.25}$ & 7.00$^{+0.30}_{-0.05}$ &  0.148 & 0.6 & 0.09$^{+0.13}_{-0.04}$ \\
         &  16 &  8 19 17.0 & +70 42 40 &  34.0 & -10.72$\pm$0.004 & 3.50$^{-.08}_{-0.26}$ & 6.48$^{+0.12}_{-0.00}$ & -0.060 & 0.4 & 0.26$^{+0.04}_{-0.06}$ \\
         &  17 &  8 19 23.8 & +70 42  8 &  34.6 &  -9.64$\pm$0.003 & 3.22$^{-.10}_{-0.30}$ & 6.30$^{+0.30}_{-0.30}$ & -0.355 & 0.9 & 0.05$^{+0.06}_{-0.04}$ \\
DDO 53   &   1 &  8 34  6.9 & +66 10 56 &  86.5 & -12.61$\pm$0.001 & 4.21$^{-.08}_{-0.26}$ & 6.48$^{+0.12}_{-0.18}$ & -0.161 & 1.0 & 0.13$^{+0.04}_{-0.06}$ \\
         &   2 &  8 34  9.8 & +66 10 44 &  42.9 &  -9.44$\pm$0.004 & 3.55$^{-.07}_{-0.15}$ & 7.18$^{+0.00}_{-0.03}$ & -0.212 & 0.6 & 0.00$^{+0.02}_{-0.00}$ \\
         &   3 &  8 34  7.8 & +66 10 51 &  37.3 & -10.39$\pm$0.003 & 3.50$^{-.10}_{-0.26}$ & 6.00$^{+0.60}_{-0.00}$ & -0.140 & 0.9 & 0.10$^{+0.05}_{-0.04}$ \\
         &   4 &  8 34  8.7 & +66 10 52 &  19.7 &  -9.70$\pm$0.003 & 3.17$^{-.08}_{-0.28}$ & 6.30$^{+0.30}_{-0.30}$ &  0.084 & 0.9 & 0.03$^{+0.05}_{-0.03}$ \\
         &   5 &  8 34  8.4 & +66 10 49 &  22.5 &  -9.74$\pm$0.004 & 3.07$^{-.08}_{-0.26}$ & 6.48$^{+0.12}_{-0.00}$ & -0.132 & 1.0 & 0.08$^{+0.04}_{-0.06}$ \\
         &   6 &  8 34  8.7 & +66 10 38 &  30.2 &  -8.69$\pm$0.006 & 2.72$^{-.06}_{-0.27}$ & 6.70$^{+0.00}_{-0.22}$ & -0.738 & 0.2 & 0.01$^{+0.04}_{-0.01}$ \\
         &   7 &  8 34  3.8 & +66 10 37 &  31.6 & -10.03$\pm$0.003 & 3.28$^{-.06}_{-0.25}$ & 6.70$^{+0.00}_{-0.22}$ & -0.218 & 0.6 & 0.07$^{+0.04}_{-0.04}$ \\
         &   8 &  8 34  8.6 & +66 10 47 &  23.2 &  -7.42$\pm$0.011 & 2.74$^{-.03}_{-0.15}$ & 7.18$^{+0.00}_{-0.03}$ & -0.489 & 0.3 & 0.00$^{+0.04}_{-0.00}$ \\
         &   9 &  8 34  3.4 & +66 10 41 &  42.2 & -10.52$\pm$0.003 & 3.47$^{+0.16}_{-0.30}$ & 6.70$^{+0.48}_{-0.22}$ & -0.278 & 0.5 & 0.11$^{+0.04}_{-0.11}$ \\
         &  10 &  8 34  9.7 & +66 10 38 &  33.8 & -10.50$\pm$0.004 & 3.77$^{-.01}_{-0.22}$ & 7.00$^{+0.04}_{-0.00}$ &  0.216 & 0.1 & 0.18$^{+0.04}_{-0.05}$ \\
         &  11 &  8 34  6.0 & +66 10 21 &  33.8 &  -9.34$\pm$0.004 & 3.01$^{-.06}_{-0.21}$ & 6.70$^{+0.00}_{-0.00}$ & -0.544 & 0.8 & 0.03$^{+0.04}_{-0.03}$ \\
DDO 63   &   1 &  9 40 45.1 & +71 11  0 & 237.9 & -12.05$\pm$0.002 & 4.42$^{+0.37}_{-0.16}$ & 7.00$^{+0.30}_{-0.00}$ & -0.830 & 0.4 & 0.02$^{+0.12}_{-0.02}$ \\
         &   2 &  9 40 34.0 & +71 09 58 & 304.4 & -12.46$\pm$0.001 & 4.59$^{+0.37}_{-0.12}$ & 7.00$^{+0.30}_{-0.52}$ & -0.874 & 0.2 & 0.00$^{+0.24}_{-0.00}$ \\
         &   3 &  9 40 39.5 & +71 10 13 & 109.4 & -11.86$\pm$0.003 & 4.05$^{-.08}_{-0.25}$ & 6.00$^{+0.30}_{-0.00}$ & -0.525 & 1.0 & 0.14$^{+0.05}_{-0.04}$ \\
         &   4 &  9 40 24.7 & +71 10 25 & 131.5 & -11.76$\pm$0.003 & 4.22$^{-.03}_{-0.22}$ & 7.00$^{+0.04}_{-0.00}$ & -0.515 & 0.0 & 0.15$^{+0.04}_{-0.05}$ \\
         &   5 &  9 40 18.2 & +71 11 22 & 129.2 & -11.53$\pm$0.001 & 4.38$^{-.07}_{-0.11}$ & 7.18$^{+0.00}_{-0.00}$ & -0.340 & 0.0 & 0.00$^{+0.02}_{-0.00}$ \\
         &   6 &  9 40 35.5 & +71 10 44 & 156.0 & -12.03$\pm$0.004 & 4.29$^{+0.03}_{-0.21}$ & 7.00$^{+0.04}_{-0.05}$ & -0.594 & 0.0 & 0.20$^{+0.05}_{-0.04}$ \\
NGC 3738 &   1 & 11 35 47.4 & +54 31 33 & 260.8 & -15.44$\pm$0.000 & 6.23$^{+0.46}_{-0.21}$ & 6.95$^{+0.74}_{-0.26}$ &  0.900 & 0.9 & 0.01$^{+0.39}_{--.15}$ \\
         &   2 & 11 35 48.4 & +54 31 18 & 138.4 & -13.77$\pm$0.001 & 6.12$^{+0.06}_{-0.71}$ & 7.70$^{+0.30}_{-0.92}$ &  1.341 & 1.0 & 0.05$^{+0.36}_{--.11}$ \\
         &   3 & 11 35 48.8 & +54 31 26 &  66.8 & -12.48$\pm$0.001 & 5.65$^{+0.07}_{-0.71}$ & 7.70$^{+0.30}_{-0.92}$ &  1.503 & 1.0 & 0.04$^{+0.37}_{--.12}$ \\
Haro 29  &   1 & 12 26 15.7 & +48 29 38 &  56.7 & -14.51$\pm$0.001 & 5.13$^{-.10}_{-0.25}$ & 6.00$^{+0.30}_{-0.00}$ &  1.126 & 0.3 & 0.12$^{+0.03}_{-0.04}$ \\
         &   2 & 12 26 16.0 & +48 29 37 &  45.3 & -13.49$\pm$0.001 & 4.76$^{-.10}_{-0.23}$ & 6.00$^{+0.30}_{-0.00}$ &  0.950 & 0.0 & 0.16$^{+0.03}_{-0.03}$ \\
         &   3 & 12 26 16.5 & +48 29 37 &  45.3 & -12.49$\pm$0.003 & 4.33$^{-.10}_{-0.25}$ & 6.00$^{+0.30}_{-0.00}$ &  0.520 & 0.8 & 0.18$^{+0.04}_{-0.03}$ \\
         &   4 & 12 26 16.3 & +48 29 40 &  57.8 & -12.57$\pm$0.002 & 4.37$^{-.10}_{-0.25}$ & 6.00$^{+0.30}_{-0.00}$ &  0.349 & 0.6 & 0.16$^{+0.04}_{-0.04}$ \\
         &   5 & 12 26 16.2 & +48 29 35 &  36.3 & -12.11$\pm$0.004 & 4.19$^{-.11}_{-0.25}$ & 6.00$^{+0.30}_{-0.00}$ &  0.574 & 0.1 & 0.24$^{+0.03}_{-0.04}$ \\
         &   6 & 12 26 17.2 & +48 29 39 &  54.4 & -12.48$\pm$0.002 & 4.20$^{-.05}_{-0.22}$ & 6.60$^{+0.00}_{-0.12}$ &  0.231 & 0.7 & 0.14$^{+0.07}_{-0.03}$ \\
         &   7 & 12 26 16.9 & +48 29 38 &  23.8 & -10.81$\pm$0.007 & 3.54$^{-.05}_{-0.24}$ & 6.60$^{+0.00}_{-0.30}$ &  0.289 & 0.9 & 0.24$^{+0.07}_{-0.04}$ \\
\enddata
\end{deluxetable}

\subsection{\it LEGUS star formation rate maps}

We also make use of SFR surface density maps produced by %\citet{sfrmaps}.
Thilker et al. (2018, in preparation).
The maps we used were made at 0.25 kpc resolution and use {\it GALEX} FUV images
to determine the SFR.
{\it Wide-field Infrared Survey Explorer (WISE)}
HiRes W4 22 $\mu$m maps, if the galaxy was detected, were used to correct for extinction.
For DDO 53 the ``unWISE'' images (unofficial, unblurred co-adds of the WISE images)
were used instead \citep{unwise1,unwise2}.
We followed the method of Jarrett \citep[private communication, see also][]{jarrett12}
%\citet[][see also Jarrett et al.\ 2012]{jarrett17} %\citep[see also][]{jarrett12} and
and the prescription by \citet{sfrmapext}, attempting to allow for the spatially variable contribution of old stellar
populations to the dust heating, with the scaling as a function of local FUV-W1 color.
This method scales the IR bands by a factor of 6.0, appropriate for local scales.
If instead we used a scaling appropriate for global measurements of galaxies, such as the value of 3.89 determined by \citet{hao11},
we find that the SFR density in spots in the centers of the galaxies would be lower by as much as a factor of two.

The FUV was corrected for foreground extinction following \citet{sfrmapforegext}.
The maps use a Kroupa IMF from 0.1 to 100 M\solar\ with a SFR timescale $\ge$100 Myr \citep{kroupaimf}.
The units of the maps are M\solar\ yr$^{-1}$ kpc$^{-2}$.
We note, however, that SFRs determined over regions, especially small regions, within dwarf galaxies can be highly affected
by the stochastic sampling of a universal IMF and time-dependent fluctuations in the SFR
\citep{fumagalli11,dasilva14}.

We have integrated the SFR in DDO 50 over all pressure regions in order to compare to the SFR determined
from the {\it GALEX} FUV based on a calibration from resolved stellar populations by \citet{mcquinn15}.
Their value, converted to our distance, is $\log {\rm SFR}=-1.21\pm0.17$. This is 1.4 times larger than our value from the
SFR map described here, but they are the same within one sigma. The integrated SFR from {\it GALEX} FUV
given in Table \ref{tab-sample} from \citet{lt} is 50\% lower than the \citet{mcquinn15} value.

\subsection{\it Galactic environments}

\begin{figure}
\epsscale{1.0}
\vskip -1.5truein
\vskip -.2truein
%\plotone{fig6_1.pdf}
\includegraphics[angle=0,width=1.0\textwidth]{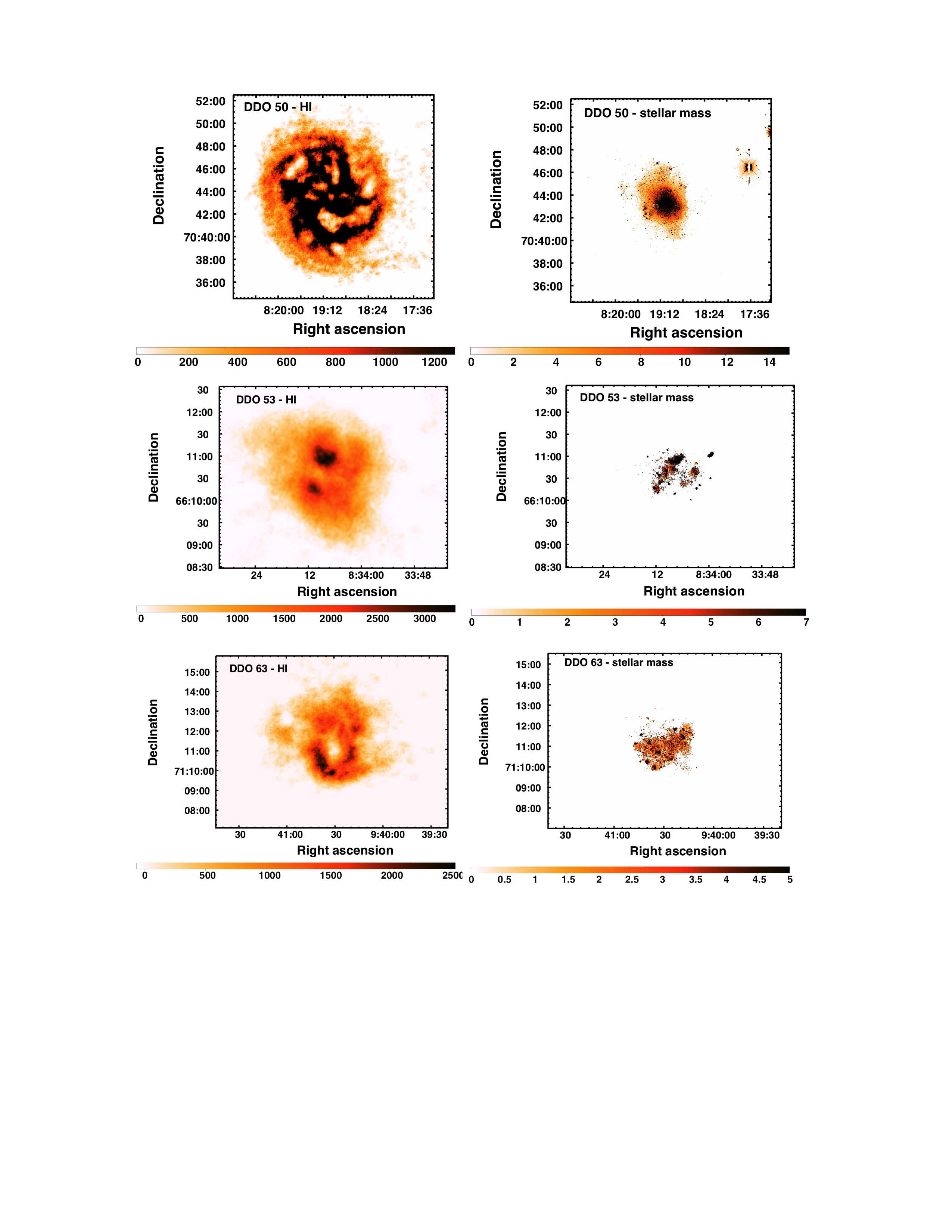}
\vskip -2.5truein
\caption{
Integrated \HI\ (moment 0) maps and stellar mass surface density maps are shown for each galaxy.
The units of the \HI\ maps are $10^{18}$ atoms cm$^{-2}$
and the units of the stellar mass surface density maps are M\solar\ pc$^{-2}$.}
\label{fig-maps}
\end{figure}

The LITTLE THINGS data sets include
\HI-line maps obtained with the Very Large Array interferometer (VLA\footnote[24]{
The VLA, now NSF's Karl G.\ Jansky Very Large Array (JVLA), is a facility of the National Radio Astronomy Observatory.
The National Radio Astronomy Observatory is a facility of the National Science Foundation
operated under cooperative agreement by Associated Universities, Inc.}).
The \HI\ maps are characterized by high sensitivity ($\leq1.1$ Jy beam$^{-1}$ per
channel), high spectral resolution ($\leq$2.6 \kms), and high angular resolution ($\sim$6\arcsec).
To obtain maps of gas surface density, we
converted the naturally-weighted integrated (moment 0) \HI\ maps to units of column density (\coldens)
and multiplied by 1.36 to account for Helium.
In addition we use $B$ and $V$ images obtained at Lowell Observatory to determine the stellar mass density
in each pixel. We used the $B-V$ color to determine the mass-to-light ratio using a formula
determined from SED fitting to the LITTLE THINGS photometry \citep{ml}, and with $L_V$ we determined the
stellar mass in each pixel. (Note that we do not use {\it WISE} NIR images to determine the stellar mass
because the dIrrs are not generally detected by {\it WISE}. We also do not use {\it Spitzer} 3.6 $\mu$m images
for S/N issues. See \citet{ml17} for a discussion on the effect of lack of red colors on stellar mass estimates).
From \citet{ml17}
we can expect that these stellar masses are good to a factor of two and we adopt an uncertainty of 0.3 dex in the
stellar mass surface densities.
The stellar mass densities were converted from a \citet{chabrierimf} stellar initial mass function to the \citet{kroupaimf} function
used in the cluster and OB association masses as indicated by \citet{ml}.
The integrated \HI\ maps and stellar mass maps are shown in Figure \ref{fig-maps}.

The \HI\ and stellar surface density maps were combined at their native resolutions to produce maps of the hydrostatic mid-plane pressure in each galaxy:
$$ P = 2.934\times10^{-55} \times \Sigma_{gas}(\Sigma_{gas} + (\sigma_g/\sigma_*)\Sigma_*)  ~~{\rm [g/(s^2 cm)],} $$
where $\Sigma$ is a surface density and $\sigma$ is a velocity dispersion \citep{bruce89}.
Here, $\Sigma_{gas}$ is determined solely from \HI$+$He since the molecular $H_2$ content is unknown.
Although the molecular gas is more closely tied to star formation than the atomic, the \HI$+$He is the
material that is available to become molecular on a larger spatial scale.
Molecular gas is most likely to be found within the denser \HI\ clouds in low metallicity environments,
so if molecular content was known and included, it would probably increase the pressure in the higher pressure regions
rather than increasing the pressure in low \HI\ density regions.
The gas velocity dispersion was derived from the \HI\ moment 2 maps.
The stellar velocity dispersion was estimated using $\log \sigma_* = -0.15M_B - 1.27$ from \citet{swaters},
where $M_B$ is the integrated absolute $B$ magnitude of the galaxy.
Because dIrrs are gas dominated and since the gas surface density enters as $\Sigma_{gas}^2$,
the pressure maps are dominated by the \HI.

We estimate the uncertainties in $\Sigma_{HI}$ in the integrated \HI\ moment 0 map
from the \HI\ data cube channel rms \citep{lt} and
assume the number of channels contributing to each pixel in the integrated moment zero map is
the typical velocity profile FWHM divided by the channel width, about 6 channels.
We take a typical pressure region of radius 200 pc and determine the number of \HI\ pixels $N$ summed
in such a region for the galaxy's distance, and the uncertainty in $\Sigma_{HI}$ goes as $\sqrt N$.
The uncertainty in $\log  \Sigma_{HI}$ is 0.02 (units M\solar\ pc$^{-2}$) for Haro 29,
0.04 for NGC 3738, 0.07 for DDO 63, 0.06 for DDO 53, and 0.09 for DDO 50.
For a typical \HI\ surface density of 10 M\solar\ pc$^{-2}$, the uncertainty in $\Sigma_{HI}$ is 5\%
for Haro 29 up to 20\% for DDO 50.
The uncertainty in the pressure is determined from the fact that the pressure
goes as $\Sigma_{HI}^2$, so the uncertainty of $\log P$ is 2$\times$ the uncertainty of $\log \Sigma_{HI}$.

For an idea of what to expect related to pressure,
Figure 2 of \citet{parr94} shows the minimum pressure expected for star formation to take place as a function of metallicity and stellar
radiation field. This predicts that most dIrr galaxies have pressures at or below the minimum.
In the sections that follow we divide the observed pressures into bins with bin 1 pressures below $4\times10^{-13}$ g s$^{-2}$ cm$^{-1}$,
bin 2 pressures up to $4\times10^{-12}$ g s$^{-2}$ cm$^{-1}$, and bin 3 pressures above that.
For context, the typical total mid-plane pressure in the solar neighborhood is of order $3\times10^{-12}$ g s$^{-2}$ cm$^{-1}$, near the
boundary between bins 2 and 3 \citep{cox05}.
The pressure in typical \HII\ regions in dIrrs is also relatively high and would fall between bins 2 and 3
but the typical disk of a dIrr is lower pressure by a factor of $\sim$10 \citep{eh00}, putting the typical dIrr disk at the boundary
between the lower pressure bins 1 and 2.
By contrast typical giant molecular clouds have much higher internal pressures, of order $4\times10^{-11}$ g s$^{-2}$ cm$^{-1}$ \citep{gmcpress},
solidly in bin 3.

Therefore, we have maps of the pressure, gas mass surface density, and stellar mass surface density
with which to characterize the environment in which a stellar object has formed.
In order to associate a particular environment with a cluster, OB association, or O star, we
divided the pressure maps into regions that sampled the different pressure environments of each galaxy.
This was done by eye from the pressure maps, and each circle is meant to roughly select a region of similar pressure
(i.e. brightness on the pressure map).
The purpose of averaging over regions is to increase the signal to noise and isolate high and low pressure areas.
The pressure maps with the regions encircled are shown in Figure \ref{fig-pres}.
All regions are shown even though many of them ended up not having clusters or O stars in them.
F275W images of all of the galaxies are shown in Figures \ref{fig-alld50} through \ref{fig-allharo29} with
the clusters, OB associations, and O stars marked along with the pressure regions.
We then measured the average pressure, gas mass surface density, and stellar mass surface density
within each of these circles. The average values associated with a given circle are assigned to the clusters, OB associations, and O stars
that fall within that circle. For those objects falling between circles, the closest circle is used.

From these figures, we note, first, that there are far more O stars than  clusters in each galaxy.
Second,  clusters are not always found where the O stars are located.
Third, O stars are often clustered, and these clusterings are what we have identified as
OB associations.

\section{Results} \label{sec-results}

\subsection{Cluster characteristics} \label{sec-cl}

\subsubsection{Characteristics as a function of galactic properties} \label{sec-gal}

In Figure \ref{fig-clenv} we plot the cluster CI and mass against their
galactic environmental properties of pressure, stellar mass density, and \HI\ surface mass density.
Each galaxy is plotted with a different symbol, but note that DDO 63 has no clusters and is included in the legend
as a reminder that it is a part of this sample.
We see that clusters are found at a wide range of pressures and \HI\ surface densities. However,
the clusters in NGC 3738 and Haro 29 are generally found at higher
pressures, stellar mass densities, and \HI\ surface densities than the clusters in the other two galaxies.
Haro 29 is likely an advanced dwarf-dwarf merger \citep{haro29} and NGC 3738 may be, too \citep{n3738}.
Perhaps such external events are necessary to produce large numbers of clusters or extraordinary star-forming regions in dwarfs.
The other three systems are more typical, likely internally-driven dIrrs \citep{lt} \citep[but see][concerning DDO 50]{bernard12,egorov17}.
%We see that there is no trend of cluster characteristics with increasing galactic environmental properties,
%so the fact that NGC 3738 and Haro 29 are further away than the other three is unlikely to be significantly affecting
%the results.
We find only one compact cluster (NGC 3738's)  in the lowest pressure range.
In the pressure range where these dwarfs form compact and likely bound stellar systems it does not show a trend of characteristics with increasing pressure.

To check whether a correlation between cluster characteristics with galactic environmental properties was being
lost in noisy cluster data, we produced Figure \ref{fig-clenv} for only cluster class 1 objects. These are clusters
that are compact and bright and less ambiguous in their classification as a compact cluster than other objects.
The numbers of clusters drops, but %: two in DDO 50, three in DDO 53, 21 in NGC 3738, zero in Haro 29.
there is no trend with this subset of clusters.

Figure \ref{fig-clenv}  is for all compact clusters with ages up to 100 Myr. In
Figure \ref{fig-clenv10} we plot the same quantities but only for clusters with ages up to 10 Myr.
There are fewer clusters (see Table \ref{tab-numbers}), but again, no trend of cluster properties
with environmental properties is seen.

Within the statistical uncertainties, the clusters cover the same range of properties independent
of the galaxy or part of the galaxy in which they are found.
This could imply that larger scale effects are more important in determining the cluster characteristics,
as proposed by the model of \citet{whitmore07} \citep[see also][]{whitmore17}.
Another possibility is that once the conditions for clustered star formation are reached,
the gravitational collapse and the fragmentation properties of the ISM drive the final star formation efficiency within the regions,
thus making cluster formation a local and stochastic process \citep[e.g.,][]{longmore14}.
Current studies of the initial cluster mass function show that it is described by a power law function with slope close to $-2$,
consistent with the hierarchical fragmentation caused by the scale-free action of turbulence.
Star formation in dIrr galaxies and BCDs is more sporadic than it is in spiral systems.
Thus the lack of a correlation may be caused by small number statistics in sampling the cluster mass function.
The lack of dependence of cluster size on the mid-plane pressure could be explained if dynamical stellar processes,
within the gravitationally bound regions where clusters formed, operate on very short time scales \citep[e.g.,][]{grudic17},
canceling any memory of the initial conditions.
This fast dynamical evolution, which includes stellar feedback, merging of sub-clumps in the young cluster,
and tidal stripping by the host galaxy,
would explain why cluster sizes and surface brightness profiles do not depend on galactic environment,
cluster age, and galaxy type \citep[e.g.,][]{grudic17,ryon17}.

One issue in comparing cluster characteristics with the cluster's environment
is determining the true environment in which the cluster formed. We restrict ourselves to clusters with
ages less than 100 Myr in order to minimize to some extent the amount by which the environment has changed since
cluster formation, although the formation of the cluster itself modifies the natal environment.
In addition, we minimize the degree to which the clusters have dissolved \citep{lamers09,baumgardt13}.

Above we chose to capture the environmental characteristics in regions defined by sampling of pressure maps.
For comparison, we also determine the environment
%1) circles centered on each cluster with increasing radius, minus
%the cluster itself, and 2)
in annuli centered on each cluster with increasing distance from the cluster.
We want to see if there is any scale at which a correlation becomes apparent.
We define a region around each cluster to be eliminated from the environmental measurements as a circle with
a radius of 25 pc so that we are not including the cluster or OB association itself.
We then measure the average pressure, gas mass surface density, and stellar mass surface density
in circles  of radii 25 pc to 150 pc in steps of 25 pc,
150 pc to 300 pc in steps of 50 pc, and 300 pc to 1 kpc in steps of 100 pc, for a total of 15 circles.
%We saw no correlation appear for any circle size. However, the area included in the larger circles is quite large,
%and, thus, potentially includes a large variety of properties.
Annuli were constructed as the area between two successive circles.

We make plots like Figure \ref{fig-clenv} for each environmental annulus and constructed an animated ensemble of the plots
in order to easily see the changes with radius.
The movie is available in the on-line materials associated with this paper.
In Figure \ref{fig-ann1} we show the same panels as in Figure \ref{fig-clenv}, but for the smallest annulus, $\sim$38 pc radius.
This explores the environment immediately around the clusters, which can contain other star forming units.
In Figure \ref{fig-movie} we also plot the cluster characteristics against pressure for annuli 1, 7, and 15 (radii of 38 pc, 225 pc, 950 pc)
as an illustration of the entire range of radii.
We find that the environmental characteristics change with annulus, steadily becoming lower in value with increasing radius.
Not only is each annulus further from the cluster with increasing radius, but the area over which the galactic characteristics are
measured increases with radius. From the first annulus to the last there is a factor of 100 increase in area. Thus,
the lower values of the larger radii annuli are likely due to averaging over a larger area, essentially smoothing out
peaks and valleys.
Nevertheless, there is no radius at which a trend of cluster characteristics with environment develops.

In Figure \ref{fig-annvsreg} we also compare for each cluster the environmental characteristics measured in the regions
shown in Figure \ref{fig-pres} with the environmental characteristics measured in the annulus immediately around
the cluster. There is a one-to-one relationship with scatter.
Since we are after the environmental parameters in which the cloud formed that formed the clusters,
we prefer the regional characteristics that provide a reasonable average over conditions
rather than characteristics determined in the close-in annulus that is subject to local variations and crowding of other
recent star formation.

\subsubsection{Characteristics by pressure region} \label{sec-reg}

\begin{figure}
\epsscale{1.1}
\vskip -1.25truein
%\plotone{fig7.pdf}
\includegraphics[angle=0,width=1.0\textwidth]{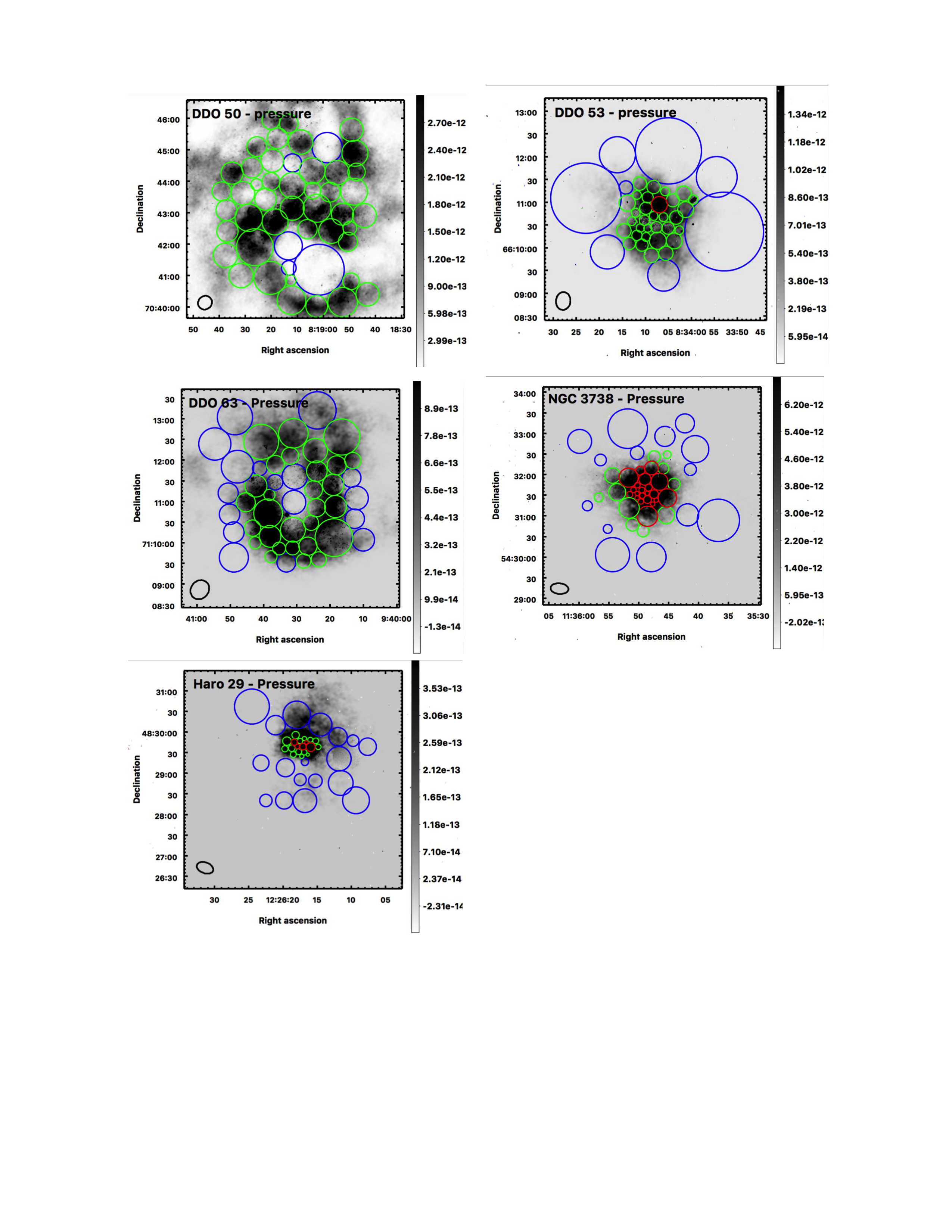}
\vskip -2.25truein
\caption{Pressure maps of the 5 galaxies showing the regions selected to sample the different pressure environment within the galaxy.
The regions and their sizes were determined by eye from the pressure maps.
The red circles are for regions with average pressures $\log P \ge -11.4$, green circles for average pressures $-12.4 \ge \log P < -11.4$,
and blue circles for average pressures $\log P < -12.4$, where the units of $P$ are g (s$^2$ cm)$^{-1}$.
The images are displayed to show structure within the inner parts of the galaxy although gas extends much further than is
obvious in these images.
The beam sizes of the \HI\ maps that form the dominant component of the pressure maps are shown as black ellipses in the lower left corner
of each panel.}
%13.7\arcsec\ (202 pc), 11.75\arcsec\ (209 pc), 14.67\arcsec\ (283 pc), 13.05\arcsec\ (310 pc), and 12.4\arcsec\ (355 pc) for
%DDO 50, DDO 53, DDO 63, NGC 3738, and Haro 29, respectively.
\label{fig-pres}
\end{figure}

\begin{figure}
\epsscale{1.0}
%\plotone{fig8.eps}
\includegraphics[angle=0,width=1.0\textwidth]{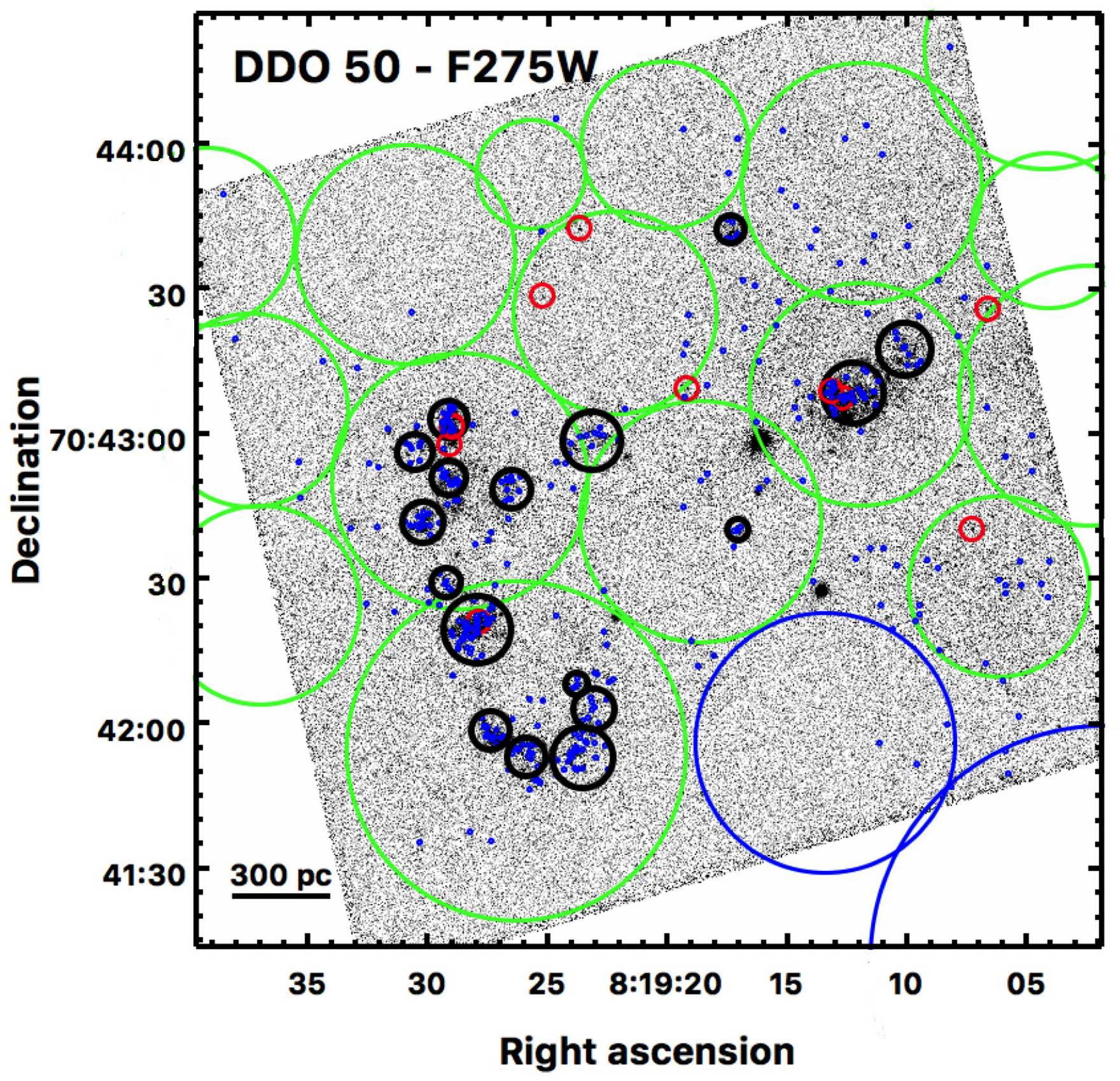}
\caption{F275W image of DDO 50 showing the regions selected to sample the pressure environment within the galaxy,
young compact clusters (small red circles), the O star candidates (small blue circles), and OB associations (black).
The large green circles are for average pressures $-12.4 \ge \log P < -11.4$ and
and blue circles for average pressures $\log P < -12.4$, where the units of $P$ are g (s$^2$ cm)$^{-1}$.
There are no regions with $\log P \ge -11.4$.
The pressure regions were determined by eye from the pressure maps;
stars or clusters between pressure regions were assigned to the closest region.
Clusters are nearly point-like and the red circles do not represent the size of the cluster.
The black circles indicate the size of the OB associations, determined by eye.}
\label{fig-alld50}
\end{figure}

\begin{figure}
\epsscale{1.0}
%\plotone{fig9.eps}
\includegraphics[angle=0,width=1.0\textwidth]{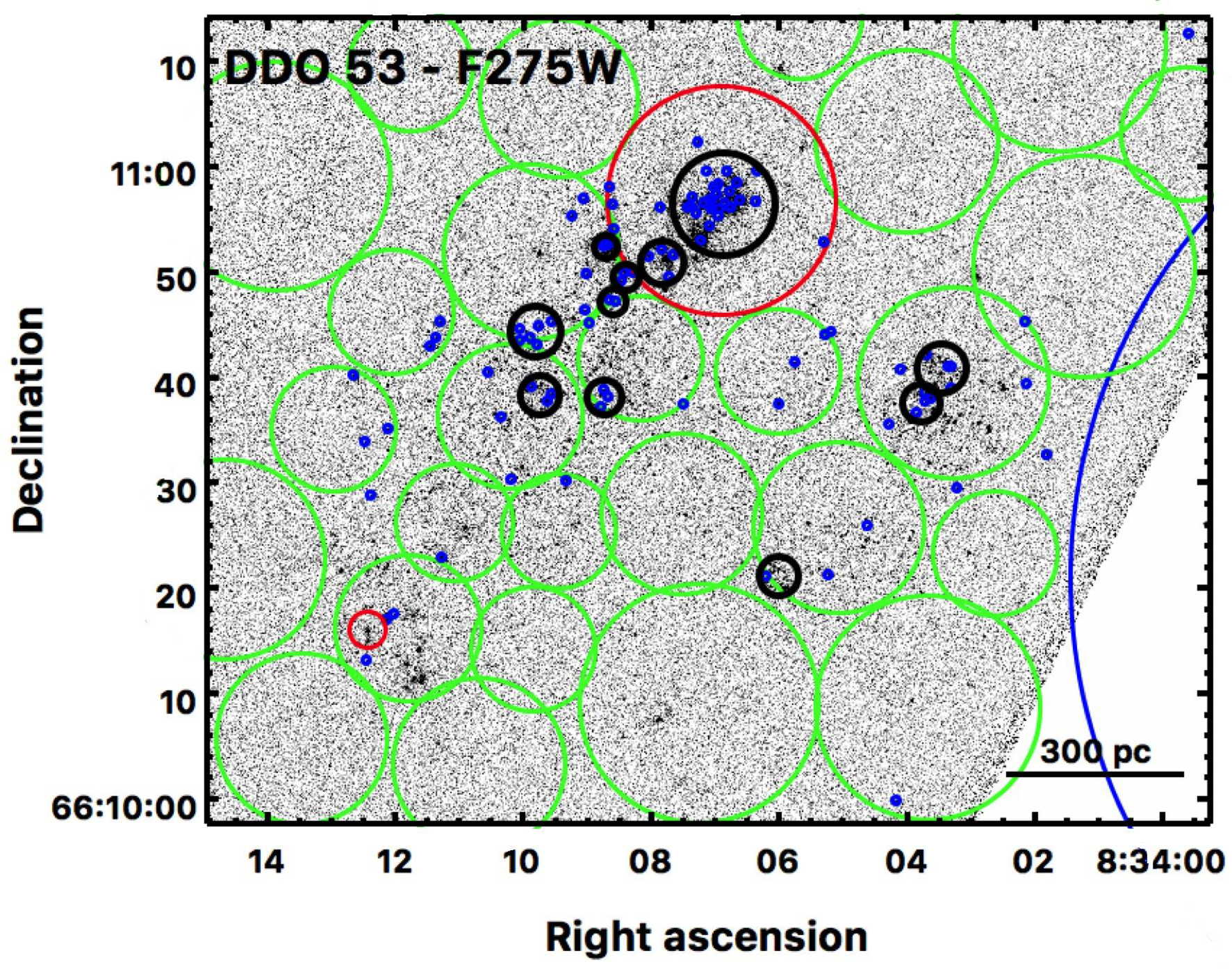}
\caption{F275W image of DDO 53 showing the regions selected to sample the pressure environment within the galaxy,
young compact clusters (small red circles), the O star candidates (small blue circles), and OB associations (black).
The large red circles are for regions with average pressures $\log P \ge -11.4$, green circles for average pressures $-12.4 \ge \log P < -11.4$,
and blue circles for average pressures $\log P < -12.4$, where the units of $P$ are g (s$^2$ cm)$^{-1}$.
Clusters are nearly point-like and the red circles do not represent the size of the cluster.
The black circles indicate the size of the OB associations, determined by eye.}
\label{fig-alld53}
\end{figure}

\begin{figure}
\epsscale{1.0}
%\plotone{fig10.eps}
\includegraphics[angle=0,width=1.0\textwidth]{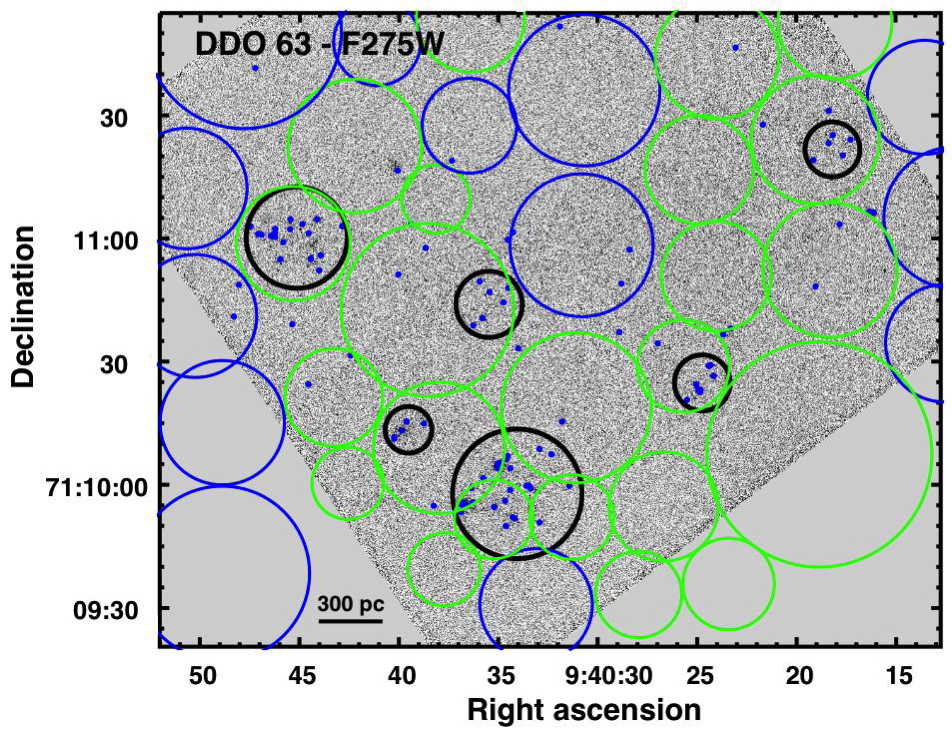}
\caption{F275W image of DDO 63 showing the regions selected to sample the pressure environment within the galaxy,
O star candidates (small blue circles), and OB associations (black).
The large green circles are for average pressures $-12.4 \ge \log P < -11.4$,
and blue circles for average pressures $\log P < -12.4$, where the units of $P$ are g (s$^2$ cm)$^{-1}$.
There are no regions with $\log P \ge -11.4$ within this galaxy.
The black circles indicate the size of the OB associations, determined by eye.
DDO 63 has no  clusters.}
\label{fig-alld63}
\end{figure}

To look at the role of pressure in another way, we examine the clusters of the pressure regions of Figure \ref{fig-pres}
in three bins; bin 1 is $\log$ pressure $<-12.4$,
bin 2 is $\log$ pressure between $-12.4$ and $-11.4$, and bin 3 is $\log$ pressure $>-11.4$,
where units of pressure are g (s$^{2}$ cm)$^{-1}$.
These bins were chosen by eye based on groupings of clusters and are marked with vertical dashed lines in Figures \ref{fig-clenv}-\ref{fig-movie}.
Note that we are only including the regions outlined in Figure \ref{fig-pres}.
These regions were chosen primarily to cover the parts of the galaxies where the stars and clusters are located, mostly the central regions,
and they do not include all of the gas associated with each galaxy, particularly extended low density gas.
Furthermore, the pressure by which the region is binned is the average within each region.
The purpose here is to describe and compare the identified pressure regions.

In the top panels of Figure \ref{fig-clarea} we plot the fraction of the total area covered by each pressure bin (panel a) and the
fraction of the \HI\ gas contained in each pressure bin (panel d).
We see, for example, that although the clusters in NGC 3738 and Haro 29 are found mostly in pressure bin 3,
the fraction of area occupied by these pressure regions is not high, 15-18\%.
On the other hand, in NGC 3738 bin 3 contains the majority of the gas, in contrast to what is seen in the rest of the systems.
For the three typical dIrrs (DDO 50, DDO 53, DDO 63) the majority of the gas is in pressure bin 2 and none to very little is
in bin 3.
Haro 29 is opposite to both of these trends, with the majority of its gas in bin 1, the lowest pressure.
In terms of cluster characteristics, the number of clusters (panels b and e) and total mass in clusters (panels c and f) per unit area and per unit gas mass
generally increase from pressure bin 2 to pressure bin 3 for the two galaxies with clusters in both bins 2 and 3 -
NGC 3738, and Haro 29. (DDO 63 has no clusters and DDO 50 and DDO 53 only have clusters in bin 2).
%If the further distances of NGC 3738 and Haro 29 have depressed the number of lower mass clusters or clusters
%have blended, the number of clusters and total cluster masses
%shown in Figure \ref{fig-clarea} for those galaxies would be lower than they should be.
We examine pressure bin 3 further by plotting, just for that pressure bin,
the ratio of the cluster mass to \HI\ mass against the fraction of the \HI\  mass in pressure bin 3 in Figure \ref{fig-cook}.
We see that the ratio of cluster mass to \HI\ mass is independent of the fraction of the \HI\ mass in pressure bin 3 although only two
galaxies have clusters in bin 3.

We were curious what the pressure distribution was among the larger sample of LITTLE THINGS dIrrs, and to look
at that we binned the pressure and integrated \HI\ maps on a pixel-by-pixel basis for 29 of the LITTLE THINGS galaxies.
Thus, here we include all of the gas, including low density gas in the outer parts.
Looking at the pressures on a pixel-by-pixel basis, we can investigate the full range of pressure environments in
our galaxies and without averaging out the highs and lows as is a consequence of our analysis of selected regions.
We found that 30\% (12) of the galaxies have no gas in pressure bin 3 while another 30\% have $>$3\% of their gas in this bin.
Of the 40 LITTLE THINGS galaxies, NGC 3738 has the highest percentage of all of its gas in bin 3, 26\%.
The starburst galaxies NGC 1569 and IC 10 have 14\%  and 10\% of their gas in bin 3, respectively.
The other galaxies in this study - DDO 50, DDO 53, DDO 63, and Haro 29 -- have 0.8\%, 1.9\%, 0, and 2\%, respectively.
We examined the LITTLE THINGS sample for any correlation between the integrated SFR and percentage of gas in pressure bin 3
and found none.

In Figure \ref{fig-clsfr} we compare the identified clusters to the SFR measured from the SFR maps
by pressure bin. First, panel (a) compares the total SFR per unit area in the three pressure bins,
while panel (b) compares the unnormalized SFR in the different bins.
Even though DDO 63 has no  clusters, it does have FUV emission, and so has a measured
SFR in bins 1 and 2.
We see that, generally the higher pressure bins have a higher SFR surface brightness.
In the right panel we compare the total mass in clusters divided by the SFR in the three
pressure bins. There are two galaxies with clusters in all three pressure bins: NGC 3738 and Haro 29.
In Haro 29 the mass formed in clusters divided by the SFR is approximately a constant with pressure.
For NGC 3738 the ratio is lower at middle pressures (bin 2) than at higher pressures (bin 3) and
at low pressure (bin 1) there is only one cluster so statistics are poor there.
The other two galaxies with clusters only in bin 2 have ratios that are comparable to NGC 3738's value in that bin.
However, this suggests that the increase in the
mass formed in clusters as a function of pressure shown in Figure \ref{fig-clarea}, panel (c),
is mainly a reflection of the fact that both the mass formed in clusters
and SFR are higher in higher pressure regions \citep[see also][]{blitz06}.
Figure  \ref{fig-clsfr}, right panel, is also consistent with the finding by \citet{chandar15}
that global cluster mass functions correlate with SFRs.
In other words, the sampling of the cluster mass function is a stochastic process driven by size-of-sample effects,
higher SFR enables sampling the cluster mass function at the high mass end \citep[e.g.,][]{adamobastian15}.

%but inconsistent with the idea that the
%fraction of stars forming in clusters, $\Gamma$, increases with the SFR (see next section).

\subsubsection{Cluster formation rate} \label{sec-gamma}

We have estimated $\Gamma$, the ratio of cluster formation rate to integrated SFR, for clusters in each of the three pressure bins.
This is not the individual circular pressure regions of Figure \ref{fig-pres}, but the sum of the clusters whose environmental
pressures fall into the three pressure bins defined in Section \ref{sec-reg}.
We have included only clusters from the team catalogues that have classes of 1 or 2 since these are compact clusters
and more likely than multi-peaked class 3 objects to be bound.
%We also include clusters with masses
%greater than or equal to 1000 M\solar, and ages less than or equal to 100 Myr. We do not exclude very young clusters,
%as is done in some studies.
%In bins with at least two clusters, we estimated the cluster formation rate as the total observed stellar mass
%divided by the age interval of 100 Myr.
%The uncertainty in $\Gamma$ takes into account the Poisson statistics of the numbers of clusters and the uncertainties
%associated with the cluster mass and age.
We also include clusters with masses greater than or equal to 1000 M\solar, and ages less than or equal to 100 Myr.
We extrapolate the mass in clusters with mass between 100 and 1000 M\solar\ assuming that the cluster mass function is described by a
power-law function with slope $-2$.
We do not exclude very young clusters, as  the tracer used to derive the SFR is sensitive to star formation between 1 and 100 Myr.
In bins with at least two clusters, we estimate the cluster formation rate as the total observed stellar mass divided by the age interval of 100 Myr.
The uncertainty in $\Gamma$ takes into account the Poisson statistics of the numbers of clusters and the uncertainties associated
with the cluster mass and age \citep[e.g., see][for a complete description of the method]{adamo-m83}.
%(e.g., see Adamo et al 2015 for a complete description of the method).
For pressure bins that contain less than two clusters, $\Gamma$ is not calculated.
Issues related to the inconsistency in the timescale over which $\Gamma$ and $\Sigma_{SFR}$ are determined
are discussed by %\citet{dwarfgamma}
Cook et al.\ (2018b, in preparation) particularly in relation to dwarf galaxies.

In Figure \ref{fig-gamma} we plot $\Gamma$ as a function of pressure bin (left panel) and as a function of the SFR density (right panel)
in the pressure bin where $\Gamma$ was calculated.
$\Gamma$ varies from 0.9\% to 33\% in the second pressure bin and from 4\% to 24\% in bin 3.
NGC 3738 in pressure bin 2 has a $\Gamma$ that is significantly higher for that pressure than for DDO 50.
For NGC 3738, with clusters in both pressure bins 2 and 3, $\Gamma$ does not change significantly between the two pressure bins.
Furthermore, $\Gamma$ does not show a correlation with the total SFR density measured by pressure bin, as shown in Figure \ref{fig-clsfr}, panel (c).
%%In that context, DDO 50 and Haro 29 are significantly different from the other three dIrrs in this sample.
However, our dIrrs do scatter around the sequence of $\Gamma$ vs.\ SFR density found in other galaxies
\citep[e.g.,][]{goddard10,adamo-mrk930,annibali11,ryon14,adamo-m83,limlee15,johnson16}: for the middle pressure bin (green squares) one galaxy
lies above the black curve and one lies near the curve. % within the uncertainties.
In the highest pressure bin (number 3, red squares in Figure \ref{fig-gamma}, right panel)
NGC 3738 lies near the black curve and Haro 29 lies well below.
In the same plot, we also include the measured SFR densities (arrows) of the regions of galaxies that do not have any estimate of $\Gamma$,
color-coded accordingly to pressure bin. The range of SFR densities for low pressure bins (blue arrows) reaches one magnitude lower than the
SFR density values where clusters are formed.
The SFR density range of intermediate and high pressure bins that form or do not form clusters (green and red arrows and squares)
spans about two orders of magnitudes, suggesting that cluster formation may still be highly stochastic, probably because of the episodic nature of
star-formation in dwarf galaxies.
There are other suggestions that $\Gamma$  measured globally for galaxies is constant \citep[see for example,][]{chandar15,chandar17}.
%and that using mixtures of age ranges creates trends that are not otherwise present \citep{chandar17}.
In that context, DDO 50 and Haro 29 are significantly different from NGC 3738 in this sample.
In Cook et al.\ (2018b, in preparation) %\citet{dwarfgamma}
we will present global values of $\Gamma$ calculated for these dwarfs and discuss the effects of using averaged
SFR densities derived with calibrated flux conversions and SFR densities derived using stellar counts and resolved star formation histories.
While the $\Gamma$ used in this work are estimated using an age range sensitive to the SFR tracer adopted,
resolved recent star formation histories will enables us to estimate $\Gamma$ within smaller age ranges %\citet{dwarfgamma}.
Cook et al.\ (2018b, in preparation).

\subsection{\HII\ regions} \label{sec-hii}

For many of the LITTLE THINGS galaxies, including four of the galaxies in this paper, we also have catalogues of \HII\ regions to a completeness limit of about $2\times10^{32}$ ergs s$^{-1}$ pc$^{-2}$
\citep{hiilum}.
These represent very young star-forming units, where the surrounding galaxy has not had much time to change
since the formation of the stars in the \HII\ region.
Therefore, we made the equivalent of Figure \ref{fig-clenv} for the \HII\ regions, which we present in Figure \ref{fig-hiichar}.
We characterize the \HII\ regions by the \ha\ surface brightness: the integrated \ha\ luminosity divided by the area covered by the \HII\ region
\citep[see][for details]{hiilum}.
The diameters of the \HII\ regions range from 10 pc to 500 pc.
In Figure \ref{fig-hiichar} the bottom panels show the \ha\ surface brightness plotted against galactic properties for the
galaxies in this study. The galactic properties in this figure were those measured in the regions defined on the pressure maps.
The top panel contains these four galaxies plus 25 more from the LITTLE THINGS sample plotted against galactic pressure.
In the top panel the pressure was measured in an annulus 200 pc wide located just beyond the \HII\ region.
With the larger sample, we do see a correlation: as galactic pressure increases, the \HII\ region \ha\ surface brightness also increases. There is no correlation between diameter of the \HII\ region and pressure.
We would expect the \ha\ surface brightness to be determined by the concentration of massive stars and gas.
So this suggests that higher concentrations of massive stars and gas are preferentially found in regions with higher pressure.

\subsection{OB associations} \label{sec-obassoc}

The OB associations are outlined in Figures \ref{fig-alld50} to \ref{fig-allharo29}, and their properties are
given in Table \ref{tab-obassoc}.
These objects are large and loose associations of O stars, which are distinct from the cluster catalogues'
class 3 objects that are compact associations.
One can see from the summary of total numbers given in Table \ref{tab-numbers} that DDO 50, DDO 53,
and DDO 63 have more OB associations than clusters. Even DDO 63 that has no clusters has six OB associations.
So for these galaxies, OB associations appear to be a better descriptor of the mode of star formation in these systems.
NGC 3738 and Haro 29 are different, perhaps because they are more extreme in SFR over all.
Haro 29 has a comparable number of clusters and OB associations, 9 and 7, respectively, and
NGC 3738 has 138 clusters and three OB associations.
In addition, the OB associations in NGC 3738 and four of those in Haro 29 have a higher stellar mass density than those
in the other three galaxies. OB association \#3 in NGC 3738 has a density that is 12 times higher than the highest
density region in DDO 50, DDO 53, or DDO63.
OB association \#1 in NGC 3738 is very large, encompassing half of the optical galaxy within 0.5 disk scale length radius
\citep{2galpaper}.

The equivalent of Figure \ref{fig-clenv} is shown for the OB associations in Figure \ref{fig-obassocenv}.
The OB association characteristics of stellar mass and mass surface density are plotted against
environmental characteristics of pressure and \HI\ surface density.
We see, again, that OB associations in NGC 3738 and Haro 29 are more massive and have a higher mass
surface density than those found in the other three galaxies.
This is unlikely to be a consequence of their further distances since the OB associations are all highly
resolved.
The associations in NGC 3738 and Haro 29 are
also found at higher pressure and mostly at higher \HI\ surface density.
The OB associations in DDO 50, DDO 53, and DDO 63 have similar masses and mass densities and
are all found in the middle pressure bin. The OB associations of DDO 53 and DDO 63 tend to be found at lower \HI\ densities.

\subsection{O star distributions} \label{sec-ob}

Here we turn our attention from star clusters and OB associations to individual O stars.
The distributions of the candidate O stars are shown in Figures \ref{fig-alld50} to \ref{fig-allharo29}.
Not all stars are captured in the selected OB associations.
This does not necessarily imply that some O stars have formed in isolation, although that is possible,
but could imply that our recognition of OB associations, especially
small or older associations, may be inadequate.
The stellar characteristics that we have to work with are absolute F275W magnitude $M_{F275W}$
and number of stars.

\subsubsection{Characteristics as a function of galactic characteristics} \label{sec-obgal}

In Figure \ref{fig-stenv} we plot the O star equivalent of Figure \ref{fig-clenv}:
stellar $M_{F275W}$ against the three environmental characteristics
pressure, stellar mass density, and \HI\ mass density.
First, we see that O stars are found at a wide range of pressures, including the lowest pressure bin 1 where no  clusters are found,
as we saw for the OB associations that contain most of the O stars.
%However, the statistics of the  clusters in most of the galaxies are poor.
Furthermore, the O stars are not coincident with the clusters.
This suggests that O stars in these dwarfs are preferentially formed in less compact units that are perhaps not bound.

Second, we see that, like the  clusters and OB associations, O stars at high pressure are found exclusively in NGC 3738 and Haro 29,
and O stars at high stellar mass density and gas mass density are mostly found in NGC 3738 and Haro 29.
Furthermore, most of the O stars in NGC 3738 and Haro 29 are found in
regions with high densities.

Third, like the properties of clusters and OB associations,
generally O star magnitudes cover a large range regardless of pressure or density.
One exception is that
O stars in all galaxies do not extend to the same faintness level.
However, the lower limits correlate with the distance to the galaxy:
DDO 50 (3.1 Mpc) stars extend to $M_{F275W}$ of $\sim -5$,
DDO 53 (3.7 Mpc) extend to $\sim -5.5$,
DDO 63 (4.0 Mpc) extend to $\sim -6$,
NGC 3738 (4.9 Mpc) extend to $\sim -6$,
and Haro 29 (5.9 Mpc) extend to $\sim -6.5$.
Therefore, the change in the lower limits are to some extent a distance effect, with
the more distant galaxies having brighter absolute magnitude limits.
In addition incompleteness due to higher backgrounds in the higher SFR galaxies may also play a role.
The horizontal dashed line in Figure \ref{fig-stenv} marks a stellar absolute magnitude in F275W $M_{F275W}$ of $-6.5$.
To see the effect of making an absolute magnitude lower limit cutoff to the stellar catalogues, look only at the stars
above this line.
Doing this, one can see that the stars in pressure bins 2 and 3 extend to more or less the same  upper magnitude.
However, the brightest stars are found in NGC 3738 and Haro 29 in pressure bin 3, but there are only a few of these, and they could
be due to blending at the further distances of these galaxies.
Also, at the low pressure side of the figure, bin 1, there are no stars brighter than about $-7.5$, although there are also fewer stars
in this pressure bin. Both of these effects could be due to
size-of-sample effects in that the SFR is also a function of the pressure with the higher SFR galaxies having many more O
stars than the lower SFR galaxies or regions \citep[see][]{whitmore17}.

To correct the F275W photometry of the O stars for extinction, we applied a modest constant correction for internal
extinction of $E(B-V)=0.05$ mag per galaxy. This corresponds to an $A_V$ of 0.16 mag.
This value is also fairly consistent with $E(B-V)$ of OB associations determined from SED fitting and given
in Table \ref{tab-obassoc}.
Additional extinction evenly distributed across a galaxy would only cause us to underestimate the luminosity of all O
stars in that galaxy by the same factor. However, differential extinction across the galaxy would affect the
inter-comparison of stars.
\citet{kahre18} have developed a method of mapping the extinction in LEGUS galaxies by determining the
reddening for each object from its photometry in the galaxy's stellar catalog \citep[see][]{sabbi18}.
We have looked at the extinction map and stellar extinction histogram for NGC 3738 as likely the worst
case in our sample of galaxies. NGC 3738 has dust lanes that are clearly visible in color images in a small
region to the north of the center of the galaxy, at an RA and Dec centered around 11$^h$ 35$^m$ 49.0$^s$,
54$^o$ 31\arcmin\ 34\arcsec. The extinction map indeed shows that this region has the highest
extinction, up to $A_V\sim1.2$, and this region contains of order 10 clusters and O stars used
in this study. However, a histogram of stellar extinctions shows that most stars have $A_V$ near zero
with a tail to higher $A_V$ that involves a relatively small number of stars.
Thus, while we have underestimated the absolute F275W magnitude of some of the stars in NGC 3738 by up
to 2 magnitudes, the numbers that are affected are small. Furthermore, since the region of heaviest
extinction is also at the highest pressure, these very luminous stars would only accentuate
the trends that we see in that galaxy (see Section \ref{sec-obreg}).

\subsubsection{Characteristics by pressure region} \label{sec-obreg}

In Figure \ref{fig-streg} we examine the O star characteristics by pressure bin,
including the third brightest $M_{F275W}$, number of O stars, and $M_{F275W}$ integrated over all of the O stars
in the pressure bin.
We choose to plot the third brightest star rather than the brightest in order to reduce statistical scatter
from using the single brightest star.
We see that the third brightest $M_{F275W}$ and integrated $M_{F275W}$ are generally
higher at higher pressure, but only in NGC 3738 is the number of O stars significantly higher
in the highest pressure bin.

In Figure \ref{fig-starea} we plot the O star characteristics per unit area and per unit \HI\ gas mass by pressure bin,
similar to Figure \ref{fig-clarea}.
We see that the number of O stars and integrated F275W flux per unit area and per unit \HI\ gas mass
increase with pressure.

Figure \ref{fig-stsfr} is similar to Figure \ref{fig-clsfr} but for O stars.
Here we plot the integrated F275W flux relative to the SFR by pressure bin.
We see that generally the higher the pressure, the higher the O star F275W flux per unit SFR.
Similarly, Figure \ref{fig-streg} showed that the maximum absolute F275W magnitude of the O stars
increases with pressure bin.
By contrast, in Figure \ref{fig-clsfr} Haro 29 has flat values of
integrated cluster mass divided by SFR with pressure bin.
The O star-cluster difference may imply that the mass of the most massive star increases with pressure
or that the ratio between ongoing star formation and star formation averaged over the past 100 Myr increases with pressure.

%\subsection{Comparing clusters and OB stars by pressure region} \label{sec-rat}

In Figure \ref{fig-rat}, left, we examine the relative number of clusters and O stars.
The top two panels show a histogram of each separately.
The bottom panel shows the ratio of clusters to O stars.
Here we have summed in 0.5 $\log$ pressure bins.
The large black Xs are the ratio of the sums of clusters and O stars in all 5 galaxies.
We see that the ratio increases with pressure.

Here the clusters are small compact clusters or tiny associations that may be bound while the O stars are primarily grouped into
physically larger associations. For a given stellar initial mass function, the number of O stars formed in the clusters is
just a scaling factor times the cluster mass.
Therefore, the ratio of number of clusters to number of O star candidates is related to the number of
O stars formed in clusters to the number formed in larger OB associations.
This relationship is messy because the clusters are not all the same mass and
the number of O stars in a particular association declines with time.
Nevertheless, the rise of the ratio of number of clusters to number of O star candidates with pressure
could indicate that the number of O stars formed in clusters compared to the number formed in
associations increases with pressure.
This is in spite of the fact that the relationship is dominated at the high end by NGC 3738 where
half of the central part of the galaxy is one giant OB association.

On the right side of Figure \ref{fig-rat} we plot the sum of the mass in clusters divided by the sum
of the mass in OB associations by pressure for each galaxy (for pressures that have at least one OB association).
This may be related to the amount of star formation taking place in bound systems relative to that taking place in unbound systems.
We find that the ratio of cluster mass to OB association mass is fairly low and fairly flat. % except for DDO 53.
%DDO 53 has a comparable mass in clusters compared to OB associations at log pressure -11.25 and 3 times more
%mass in clusters compared to OB associations at log pressure -11.75.
When masses are summed over all galaxies (the large black Xs in Figure  \ref{fig-rat}), the high values at log pressure
-11.25 and -10.75 are also driven by NGC 3738, which has one very large OB association that is assigned to log pressure -10.4.
%\section{Discussion} \label{sec-discuss}

\section{Summary} \label{sec-summary}

In order to examine the role environmental factors play in determining characteristics of typical star-forming units,
we present a comparison of the concentrations, masses, and formation rates of young ($\le$100 Myr)  compact star clusters,
surface brightnesses of \HII\ regions, masses and mass surface densities of large and loose OB associations,
and distributions and F275W magnitudes of candidate O stars
with surrounding galactic pressure, stellar mass density, \HI\ surface density, and SFR surface density.
Our sample consists of three dIrr galaxies and two BCD-like galaxies within 5.9 Mpc. For \HII\ region characteristics
we include an additional 25 dIrrs and BCDs from the LITTLE THINGS sample.
We find the following:

\begin{itemize}

\item The BCD galaxies are more extreme than the dIrrs: most  of  their clusters, OB associations,
and O star candidates are found at higher pressures, stellar mass
densities, and \HI\ mass densities. In addition NGC 3738 has an extraordinary OB association that
occupies half of the inner part of the galaxy \citep[see][]{2galpaper}.
Both of these galaxies may be dwarf-dwarf mergers \citep{haro29,n3738}, and perhaps such events
are necessary to produce large numbers of clusters or extraordinary regions in dwarfs.

\item There is no trend of cluster characteristics with environmental properties,
implying that larger scale effects are more important in determining cluster characteristics
\citep[e.g.,][]{whitmore07} or that rapid dynamical evolution
(such as stellar feedback, merging of sub-clumps in the young cluster, tidal stripping by the host galaxy)
 is taking place in bound stellar systems that erases memory of the initial conditions.

\item The most massive OB associations are found at higher pressure and \HI\ surface densities,
and there is a trend of higher \HII\ region \ha\ surface brightness with higher pressure, suggesting
that a higher concentration of massive stars and gas are found preferentially in regions of higher pressure.
Furthermore, the SFR per unit area increases with pressure.

\item The number of  clusters and total mass in clusters per unit area and per unit gas mass generally increase
with pressure, while the mass formed in clusters divided by the SFR is approximately a constant with pressure.

\item $\Gamma$, the ratio of cluster formation rate to SFR, does not show a correlation with the total SFR density,
but the galaxies studied here do scatter in $\Gamma$ around the sequence of $\Gamma$ vs.\ SFR density found in other galaxies.

\item O star candidates are found at a wide range of pressures, including low pressures where bound clusters and
OB associations are not found. %/, suggesting that star formation may proceed at lower efficiency in low pressure regions.

\item The number of candidate O stars and the integrated F275W flux per unit area and per unit \HI\ gas mass increase with pressure.
Furthermore, the total O star F275W flux per unit SFR and the third brightest absolute F275W magnitude of an O star
increase with pressure.
This may imply that the mass of the most massive star increases with pressure
or that the ratio between ongoing star formation and star formation averaged over the past 100 Myr increases with pressure.

\item The ratio of the number of  clusters to number of O star candidates increases with pressure, perhaps
reflecting an increase in clustering properties with star formation rate.

\end{itemize}

\acknowledgments
Results presented here are based on observations made with the NASA/ESA {\it Hubble Space Telescope}
under the LEGUS survey.
Support for {\it HST} Program number 13364 was provided by NASA through a
grant from the Space Telescope Science Institute, which is operated by the
Association of Universities for Research in Astronomy, Incorporated, under NASA
contract NAS5-26555.
A.A. acknowledges the support of the Swedish Research Council (VetenskapsrŒdet) and the Swedish National Space Board (SNSB).
S.G. appreciates funding from the National Science Foundation grant AST-1461200 to Northern Arizona University for
Research Experiences for Undergraduates summer internships
and Drs.\ Kathy Eastwood and David Trilling for running the NAU REU program in 2016.
S.G. also appreciates the support of the 2016 CAMPARE Scholar program and Dr.\ Alexander Rudolph for directing that program.

Facilities: \facility{HST(ACS,WFC3)}, \facility{VLA}, \facility{Lowell Observatory}

\newpage
%fig1

\newpage
%fig2

\newpage
%fig3

\newpage
%fig4

\newpage
%fig5

%fig6

\clearpage
%\plotone{fig6_2.pdf}
\includegraphics[angle=0,width=1.0\textwidth]{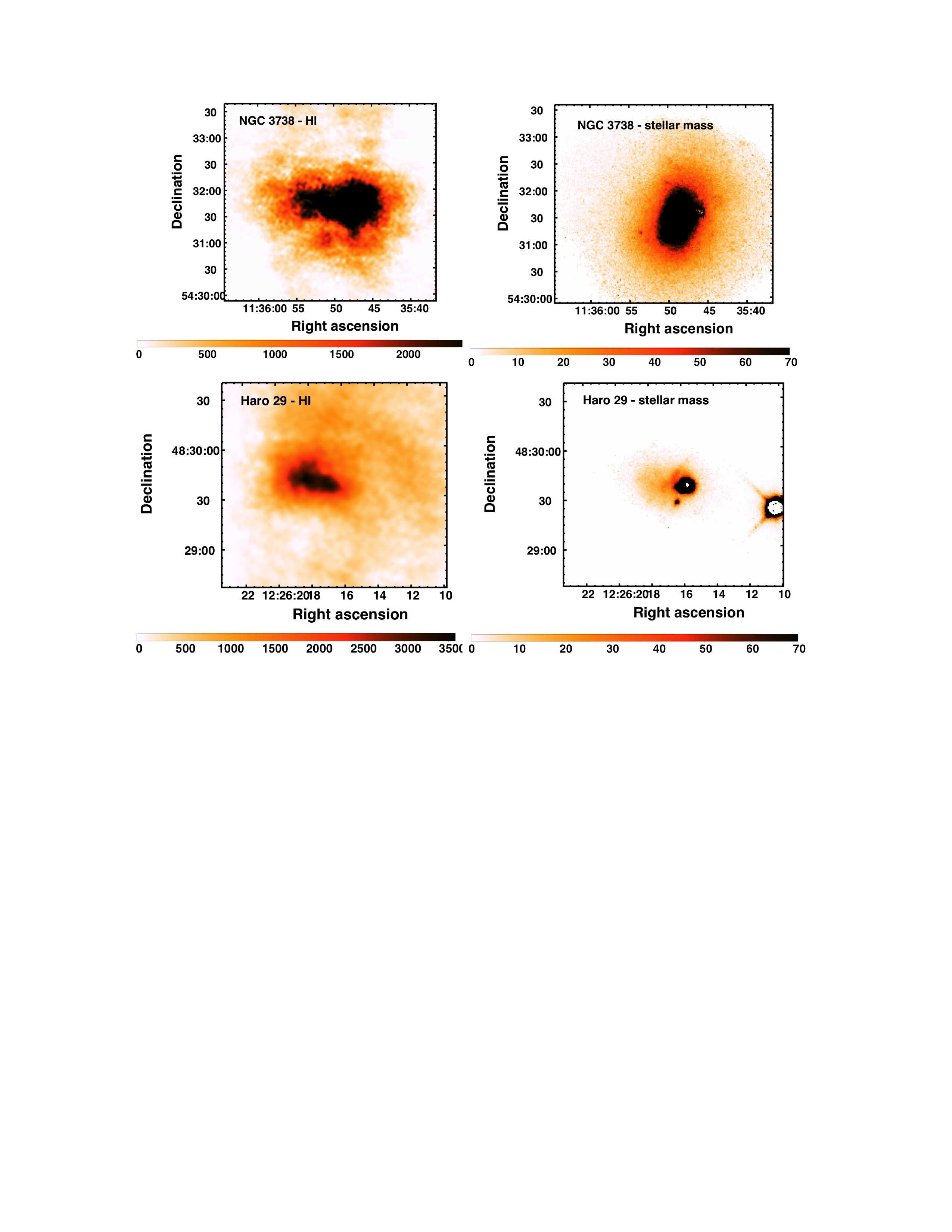}
%Figure 6 continued

\clearpage

%fig7

\newpage
%fig8

\newpage
%fig9

\newpage
%fig10

%fig11
\begin{figure}
\epsscale{1.0}
%\plotone{fig11.eps}
\includegraphics[angle=0,width=1.0\textwidth]{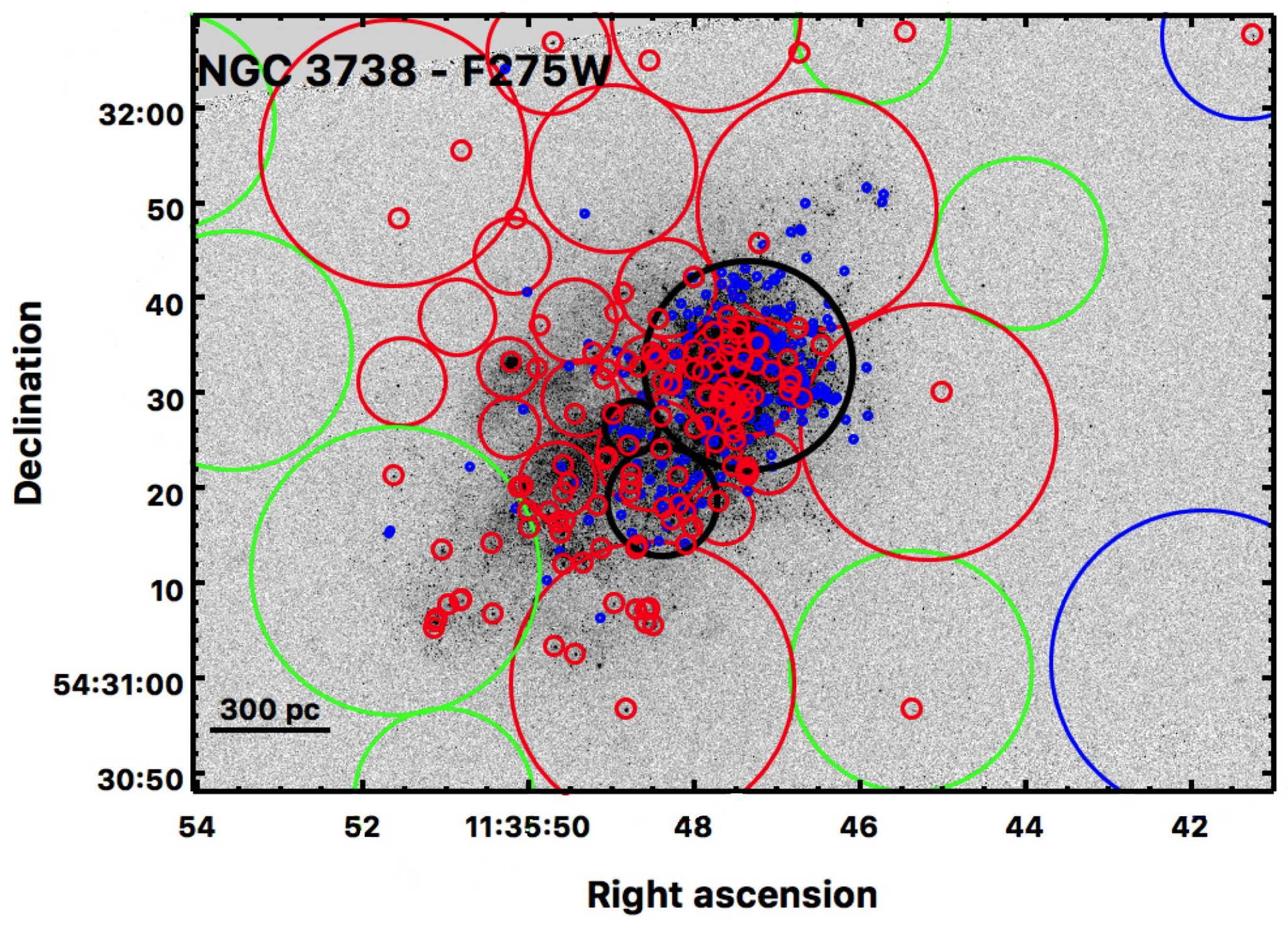}
\caption{F275W image of NGC 3738 showing the regions selected to sample the pressure environment within the galaxy,
young compact clusters (small red circles), the O star candidates (small blue circles), and OB associations (black).
The large red circles are for regions with average pressures $\log P \ge -11.4$,
green circles for average pressures $-12.4 \ge \log P < -11.4$,
and blue circles for average pressures $\log P < -12.4$, where the units of $P$ are g (s$^2$ cm)$^{-1}$.
Clusters are nearly point-like and the red circles do not represent the size of the cluster.
The black circles indicate the size of the OB associations, determined by eye.}
\label{fig-alln3738}
\end{figure}

%fig12
\begin{figure}
\epsscale{1.0}
%\plotone{fig12.eps}
\includegraphics[angle=0,width=1.0\textwidth]{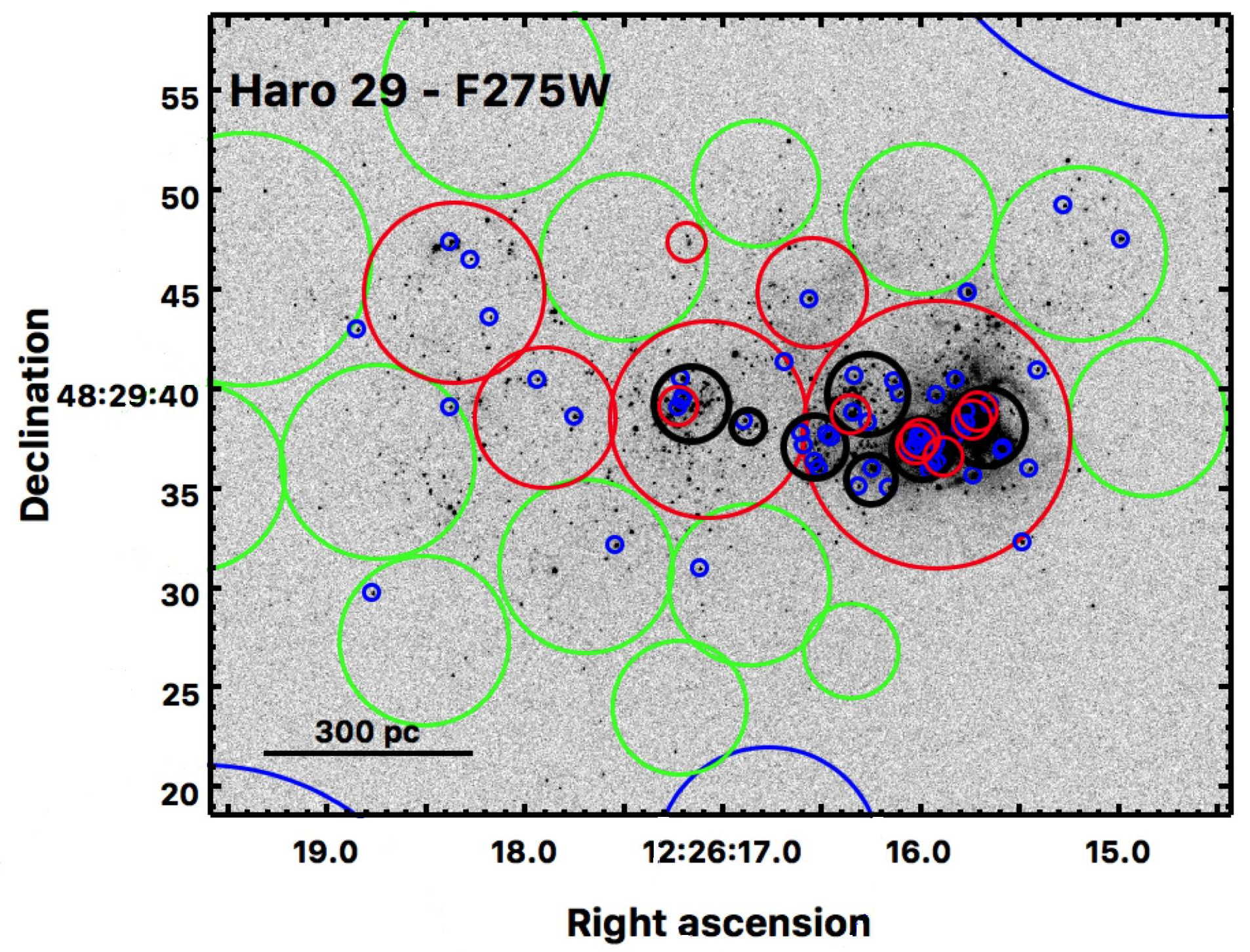}
\caption{F275W images of the Haro 29 showing the regions selected to sample the pressure environment within the galaxy,
young compact clusters (small red circles), the O star candidates (small blue circles), and OB associations (black).
The large red circles are for regions with average pressures $\log P \ge -11.4$, green circles for average pressures $-12.4 \ge \log P < -11.4$,
and blue circles for average pressures $\log P < -12.4$, where the units of $P$ are g (s$^2$ cm)$^{-1}$.
Clusters are nearly point-like and the red circles do not represent the size of the cluster.
The black circles indicate the size of the OB associations, determined by eye.}
\label{fig-allharo29}
\end{figure}

%fig13
\begin{figure}
\epsscale{1.0}
\vskip 0.25truein
%\plotone{fig13.eps}
\includegraphics[angle=0,width=1.0\textwidth]{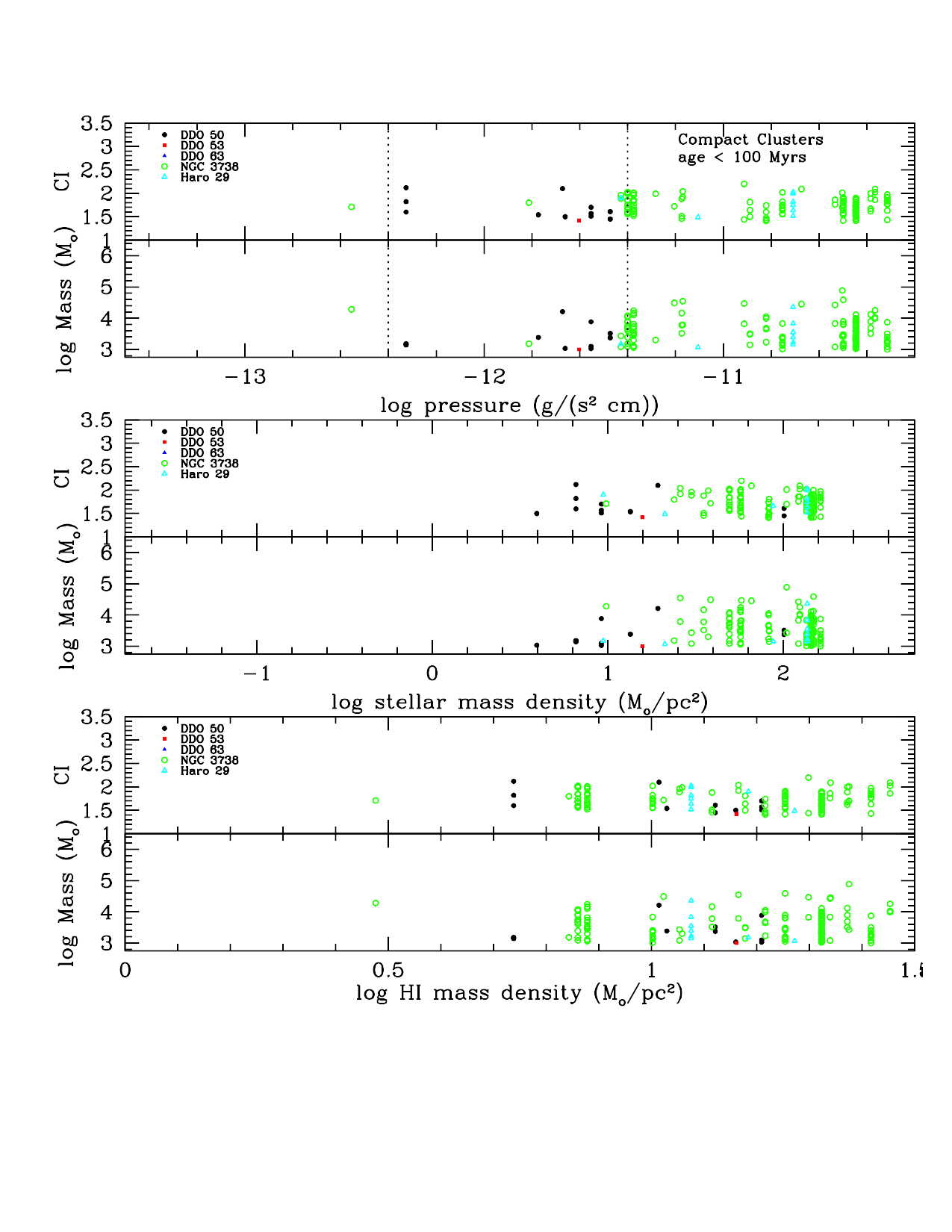}
\vskip -1.6truein
%\vskip -.25truein
\caption{Compact cluster characteristics vs.\ galactic environment in which the clusters are found
for clusters with ages less than 100 Myr.
The cluster characteristics include cluster mass and CI.
Galactic environmental characteristics include pressure, stellar mass density, and \HI\ mass surface density.
The vertical dotted lines in the top panel delineate the three pressure bins discussed in the text (Section \ref{sec-reg}).}
\label{fig-clenv}
\end{figure}

%fig14
\begin{figure}
\epsscale{1.0}
\vskip -0.25truein
%\plotone{fig14.eps}
\includegraphics[angle=0,width=1.0\textwidth]{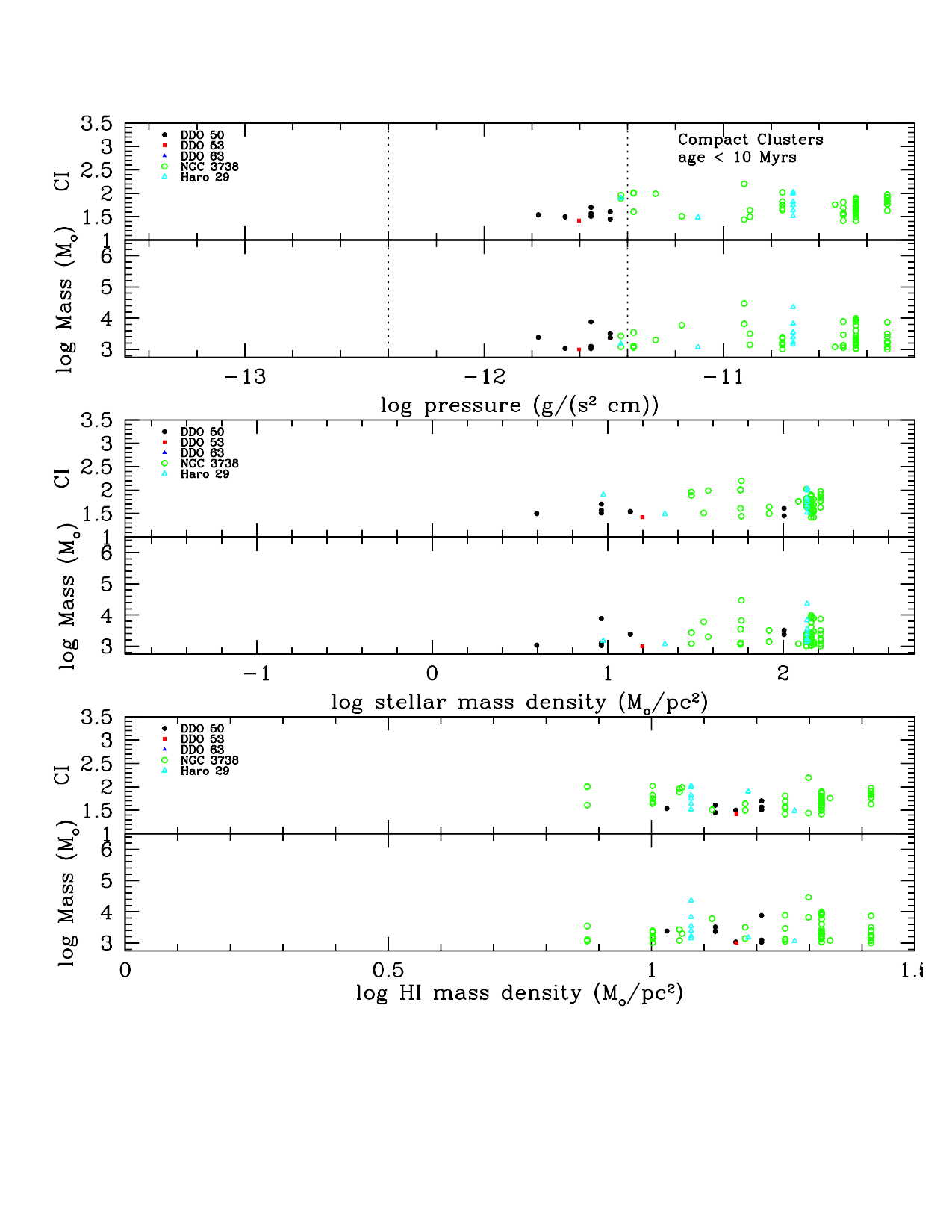}
\vskip -1.6truein
%\vskip -.25truein
\caption{Compact cluster characteristics vs.\ galactic environment in which the clusters are found
for clusters with ages less than 10 Myr.
The cluster characteristics include cluster mass and CI.
Galactic environmental characteristics include pressure, stellar mass density, and \HI\ mass surface density.
The vertical dotted lines in the top panel delineate the three pressure bins discussed in the text (Section \ref{sec-reg}).}
\label{fig-clenv10}
\end{figure}

%fig15
\begin{figure}[t!]
\epsscale{0.85}
\vskip -1.25truein
%\plotone{fig15.eps}
\includegraphics[angle=0,width=1.0\textwidth]{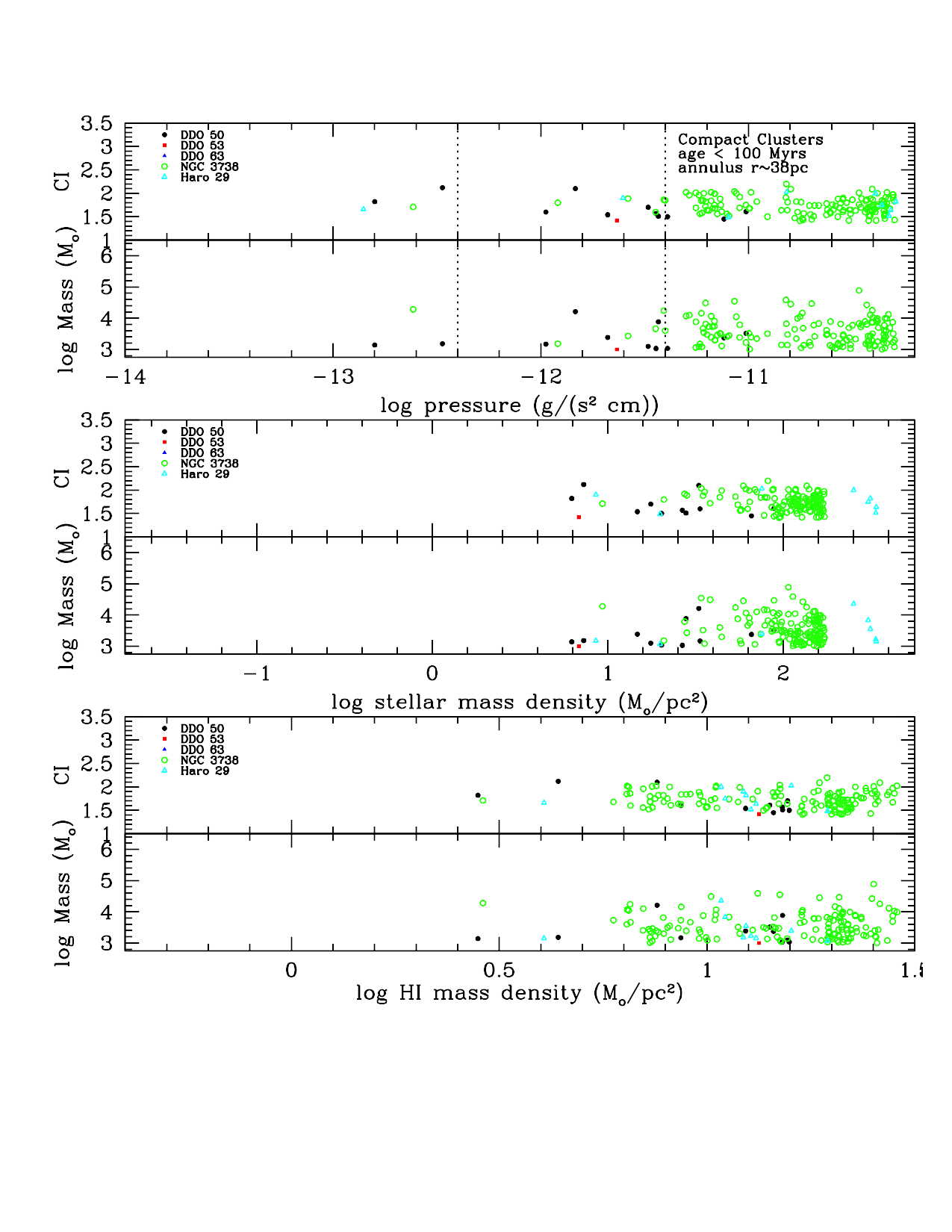}
\vskip -1.6truein
%\vskip -.25truein
\caption{Comparison of
cluster characteristics with galactic environment determined from the smallest annulus, $\sim$38 pc.
This is similar to Figure \ref{fig-clenv} but with the galactic environment determined from an annulus around the cluster
rather than selected regions.
The cluster characteristics include cluster mass and CI.
Galactic environmental characteristics include pressure, stellar mass density, and \HI\ mass surface density.
The vertical dotted lines delineate the three pressure bins discussed in the text.
The radius is the mid-point of the annulus.
There is no trend with cluster characteristic.}
\label{fig-ann1}
\end{figure}

\newpage
%fig16
\begin{figure}[h!]
\epsscale{0.85}
\vskip -1.25truein
%\plotone{fig16.eps}
\includegraphics[angle=0,width=1.0\textwidth]{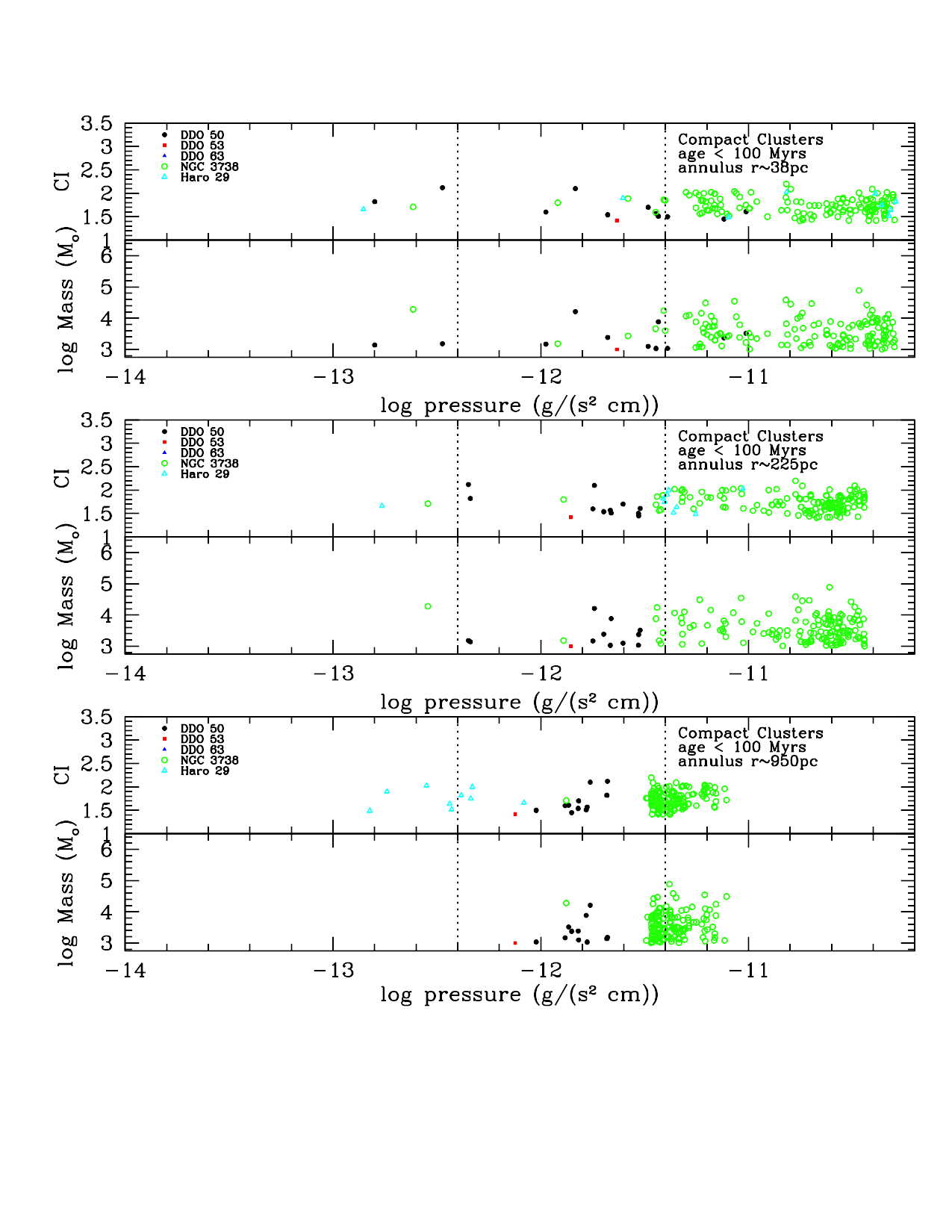}
\vskip -1.55truein
\caption{Extract of a movie comparing
cluster characteristics with galactic environment determined from annuli of progressively larger distance from the cluster.
This is similar to Figure \ref{fig-clenv} but with the galactic environment determined from annuli rather than selected regions.
The cluster characteristics include cluster mass and CI.
Galactic environmental characteristics include pressure, stellar mass density, and \HI\ mass surface density,
although here we only show the pressure panels.
The vertical dotted lines delineate the three pressure bins discussed in the text.
The movie, which is available on-line, is an animated gif with 15 annuli. Here we show the first, 7th, and last annulus.
One can see that the pressure changes with the area of the galaxy being sampled, so that clusters at one pressure in
the top panel will appear at a different pressure in the lower panel.
The radius is the mid-point of the annulus.
There is no trend with cluster characteristic at any radius.}
\label{fig-movie}
\end{figure}

\newpage
%fig17
\begin{figure}[h!]
\epsscale{0.85}
\vskip -1.25truein
%\plotone{fig17.eps}
\includegraphics[angle=0,width=1.0\textwidth]{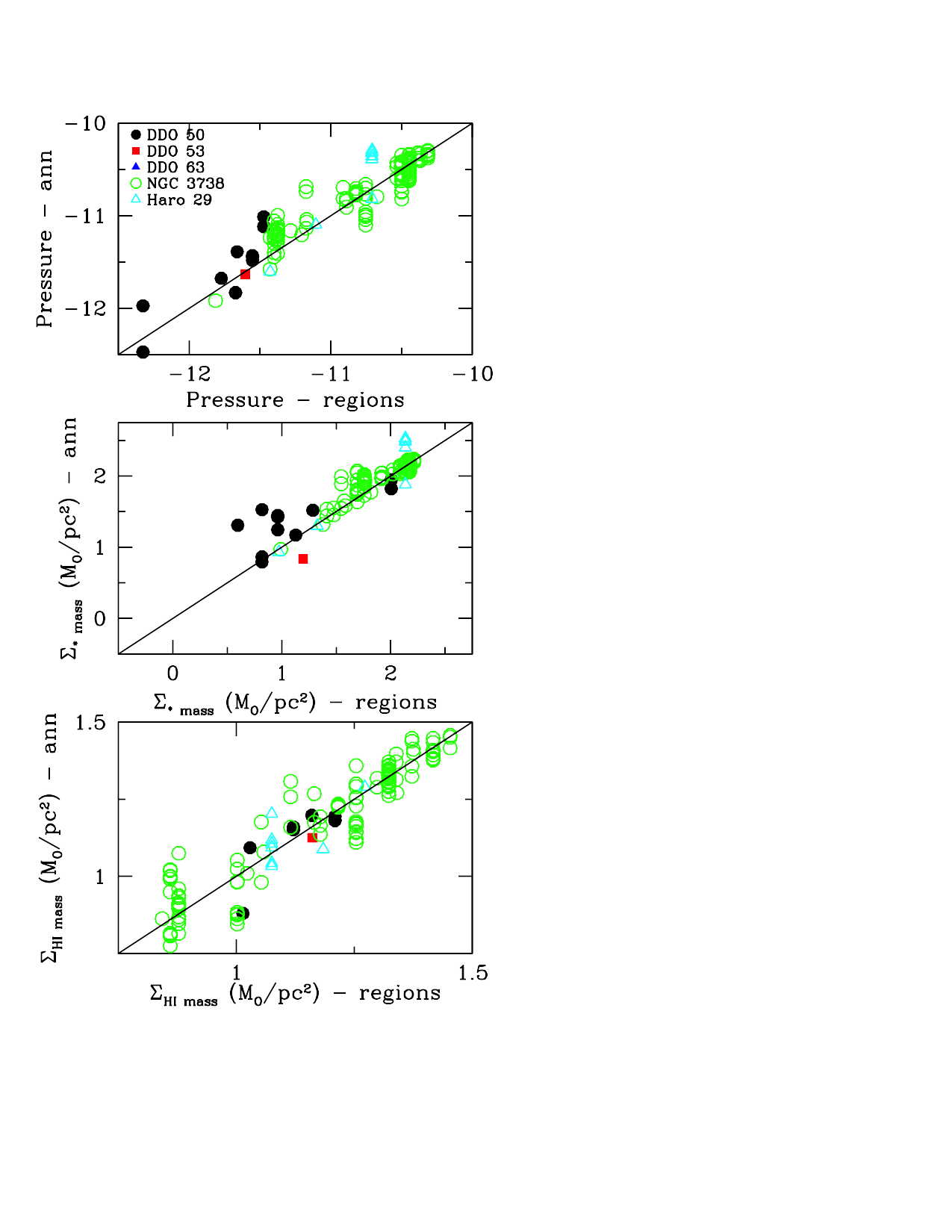}
\vskip -1.5truein
\caption{Comparison of galactic environmental characteristics determined from the smallest annulus, $\sim$38 pc,
with values determined from averages over the regions shown in Figure \ref{fig-pres} for each compact cluster.
Galactic environmental characteristics include pressure, stellar mass density, and \HI\ mass surface density.
The solid line is a one-to-one equality of the characteristics, and the relationship is one-to-one with scatter.
Since we are after the environmental parameters in which the cloud formed that formed the clusters,
we prefer the regional characteristics that provide a reasonable average over conditions
rather than characteristics determined in the close-in annulus that is subject to local variations and crowding of other
recent star formation.}
\label{fig-annvsreg}
\end{figure}

\newpage
%fig18
\begin{figure}[t!]
\epsscale{0.8}
%\vskip 1.0truein
%\plotone{fig18.eps}
\includegraphics[angle=0,width=1.0\textwidth]{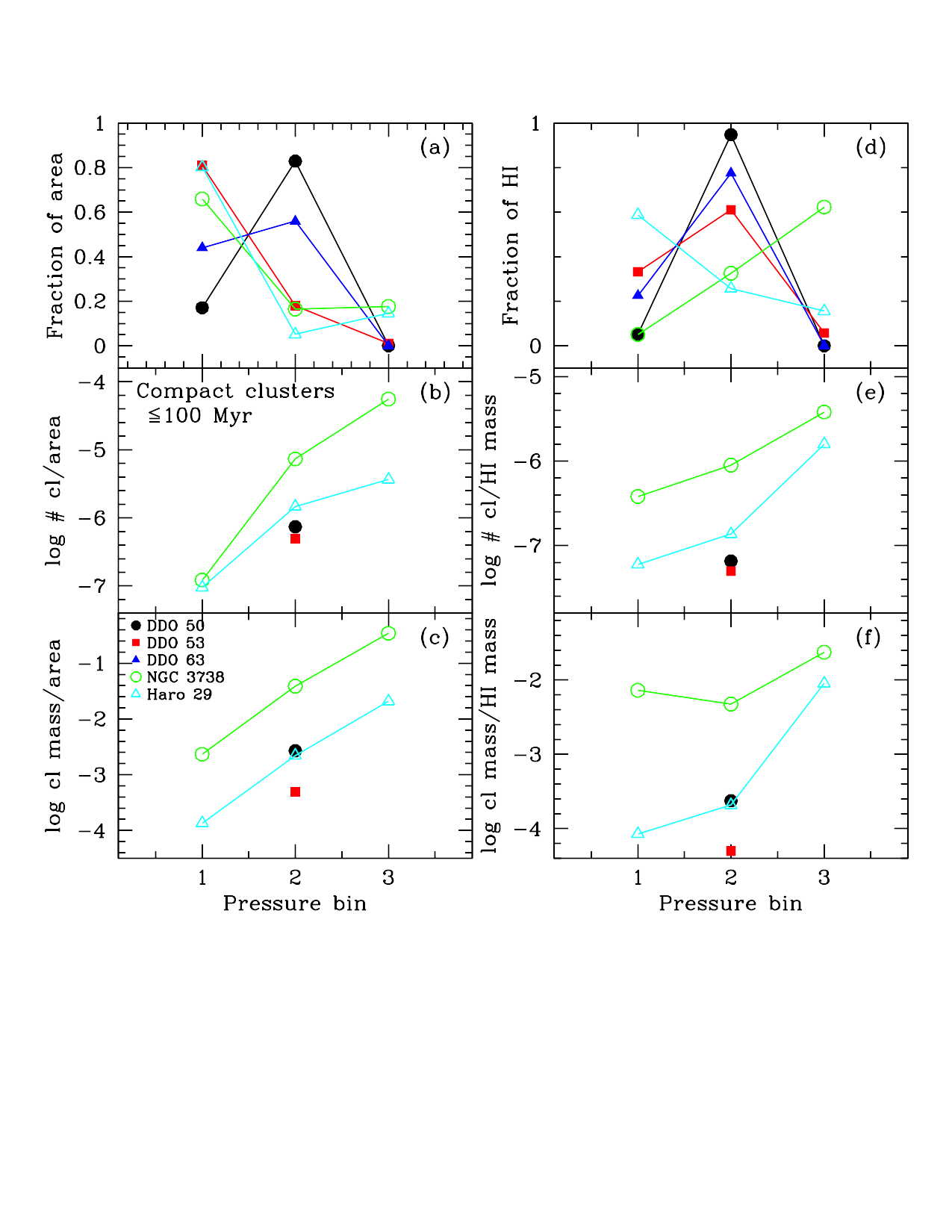}
\vskip -1.9truein
\caption{Number (panels b and e) and total mass (panels c and f) of  clusters per unit area (panels a-c) and per \HI\ gas mass (panels d-f)
vs.\ pressure in which the clusters are found.
The pressures are combined in three bins; bin 1 is $\log$ pressure $<-12.4$,
bin 2 is $\log$ pressure between $-12.4$ and $-11.4$, and bin 3 is $\log$ pressure $>-11.4$.
Units of pressure g (s$^{2}$ cm)$^{-1}$, units of area are pc$^{2}$, and the units of mass are M\solar.
The area (panel a) is the total area of pressure regions shown in Figure \ref{fig-pres} within the given bin range;
similarly for the \HI\ mass (panel d).
DDO 63 does not contain any  clusters and DDO 50 and DDO 53 do not have any clusters in pressure
bins 1 or 3.}  %None of the galaxies has clusters in pressure bin 1, the  lowest pressure.
\label{fig-clarea}
\end{figure}

\newpage
%fig19
\begin{figure}[t!]
\epsscale{0.8}
\vskip +1.0truein
%\plotone{fig19.eps}
\includegraphics[angle=0,width=1.0\textwidth]{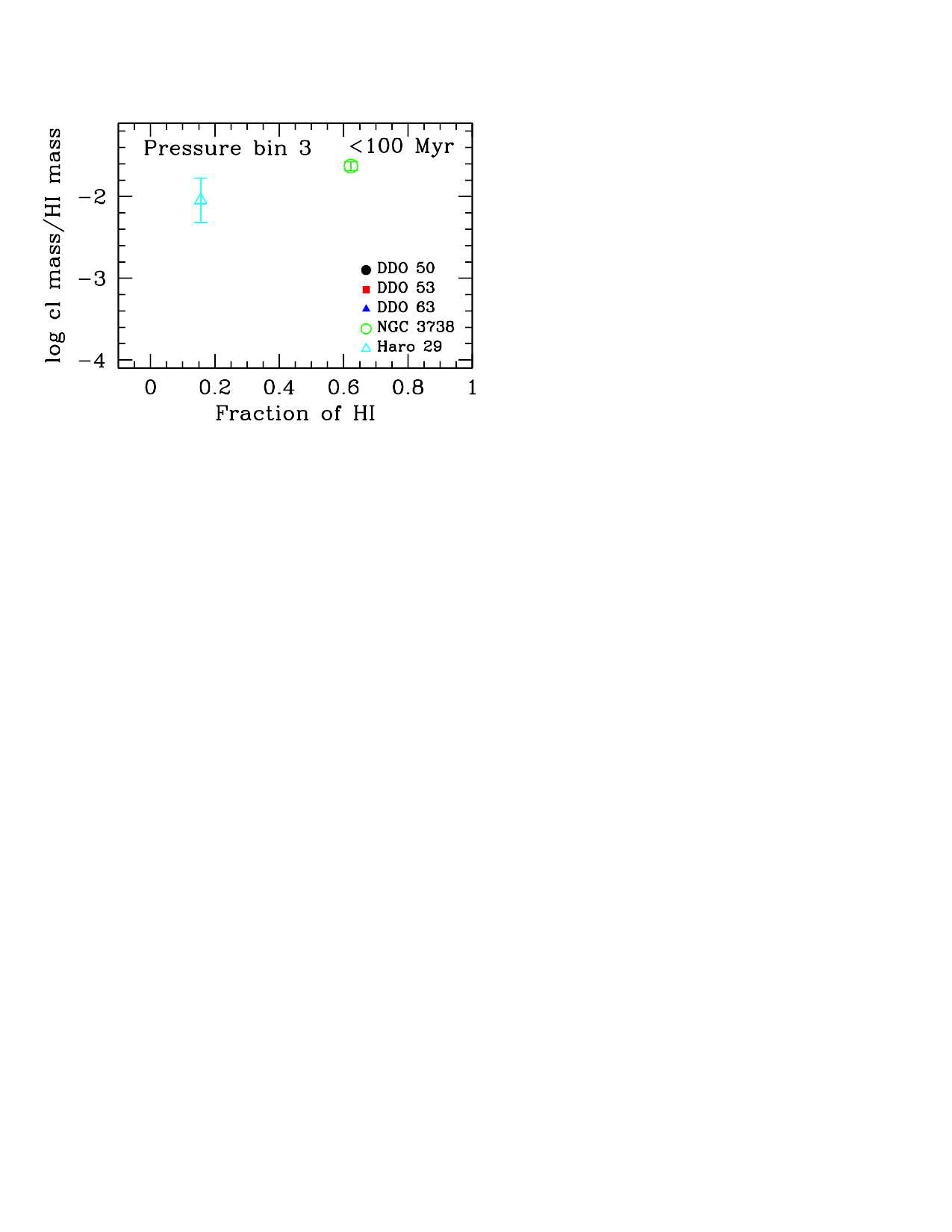}
\vskip -4.8truein
\caption{For pressure bin 3 only, total mass of clusters per \HI\ gas mass vs.\ fraction of \HI\ mass in this pressure bin.
Pressure bin 3 is $\log$ pressure $>-11.4$.
Units of pressure g (s$^{2}$ cm)$^{-1}$.
DDO 63 does not contain any  clusters and DDO 50 and DDO 53 do not have any clusters in pressure bin 3.
The ratio of cluster mass to \HI\ mass is constant for a range in fraction of \HI\ mass.}
\label{fig-cook}
\end{figure}

\newpage
%fig20
\begin{figure}[t!]
\epsscale{1.0}
%\vskip -0.5truein
%\plotone{fig20.eps}
\includegraphics[angle=0,width=1.0\textwidth]{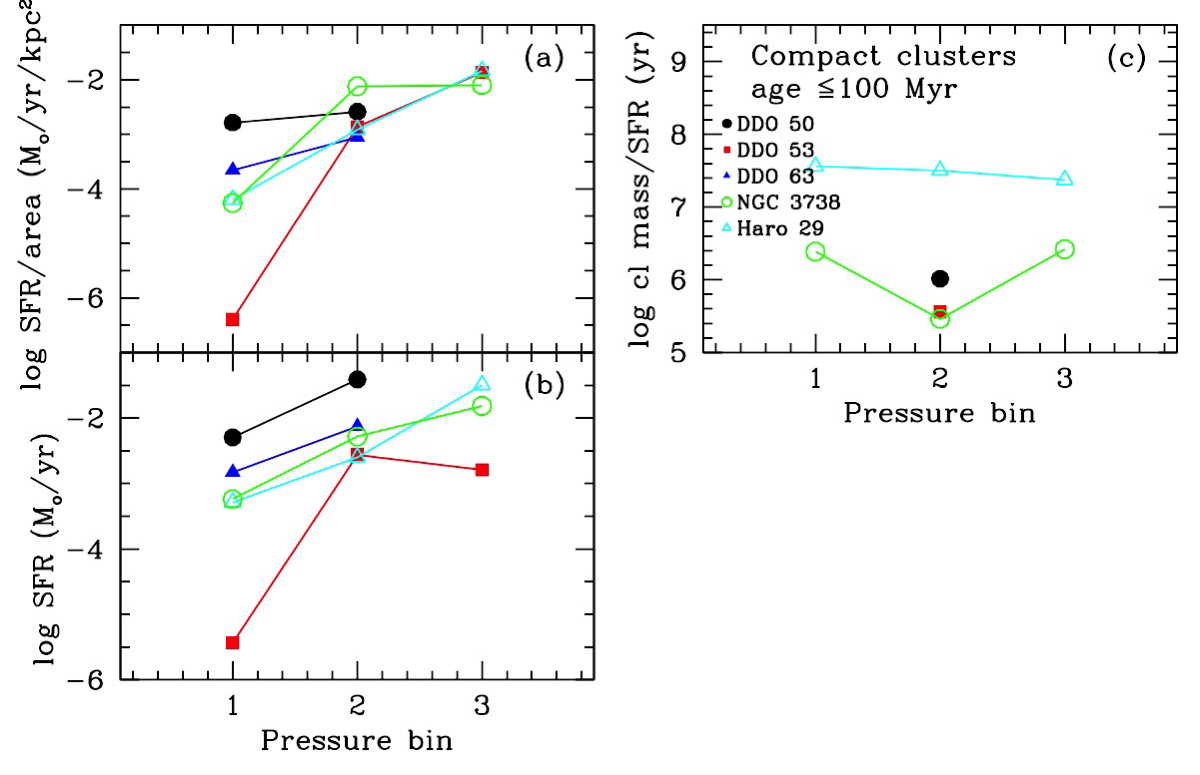}
\vskip -0.25truein
\caption{
{\it Panel (a):} SFR per unit area by pressure bin.
{\it Panel (b):} SFR by pressure bin.
{\it Panel (c):} Total cluster mass divided by the SFR by pressure bin.
The pressures within each galaxy are combined in three bins; bin 1 is $\log$ pressure $<-12.4$,
bin 2 is $\log$ pressure between $-12.4$ and $-11.4$, and bin 3 is $\log$ pressure $>-11.4$.
Units of pressure g (s$^{2}$ cm)$^{-1}$, units of SFR are M\solar\ yr$^{-1}$, and the units of mass are M\solar.
The area is the total area of pressure regions shown in Figure \ref{fig-pres} within the given bin range.
DDO 63 does not contain any  clusters and DDO 50 does not have any clusters in pressure
bins 1 or 3.  None of the galaxies has clusters in pressure bin 1.}
\label{fig-clsfr}
\end{figure}

\newpage
%fig21
\begin{figure}[t!]
%\epsscale{1.0}
\centering%
%\vskip -0.5truein
%\plotone{../angela/gamma_pressure_trim.eps}
%\plottwo{../angela/gamma_pressure_trim.eps}{../angela/gamma_dwarfs_cibased_pressurebin.eps}
%\special{psfile="fig21left.eps" hoffset=-10 voffset=-150 hscale=70 vscale=70}
%\special{psfile="fig21right.eps" hoffset=200 voffset=-200 hscale=50 vscale=50}
%%\includegraphics[angle=0,width=1.0\textwidth]{fig21left.jpg,fig21right.jpg}
%\begin{subfigure}{.5\textwidth}
%  \centering
%  %\includegraphics[width=1.0\linewidth]{fig21left.jpg}
%\end{subfigure}%
%\begin{subfigure}{.5\textwidth}
%  \centering
%  %\includegraphics[width=1.0\linewidth]{fig21right.jpg}
%\end{subfigure}
    %\subfloat[]{{\includegraphics[width=7.5cm]{fig21left.eps} }}%
    %\qquad
    %\subfloat[]{{\includegraphics[width=7.5cm]{fig21right.eps} }}%
    \includegraphics[width=7.5cm]{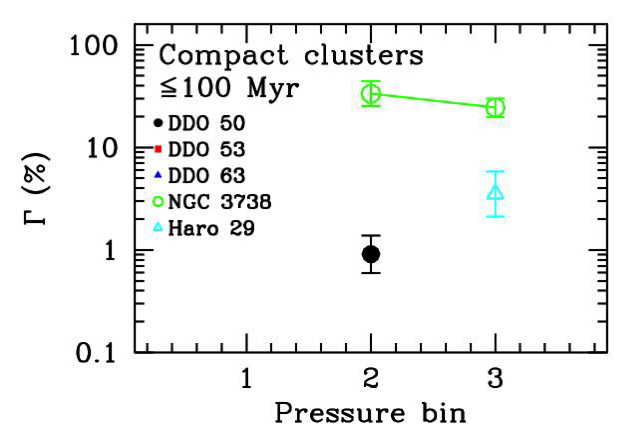}
    \includegraphics[width=7.5cm]{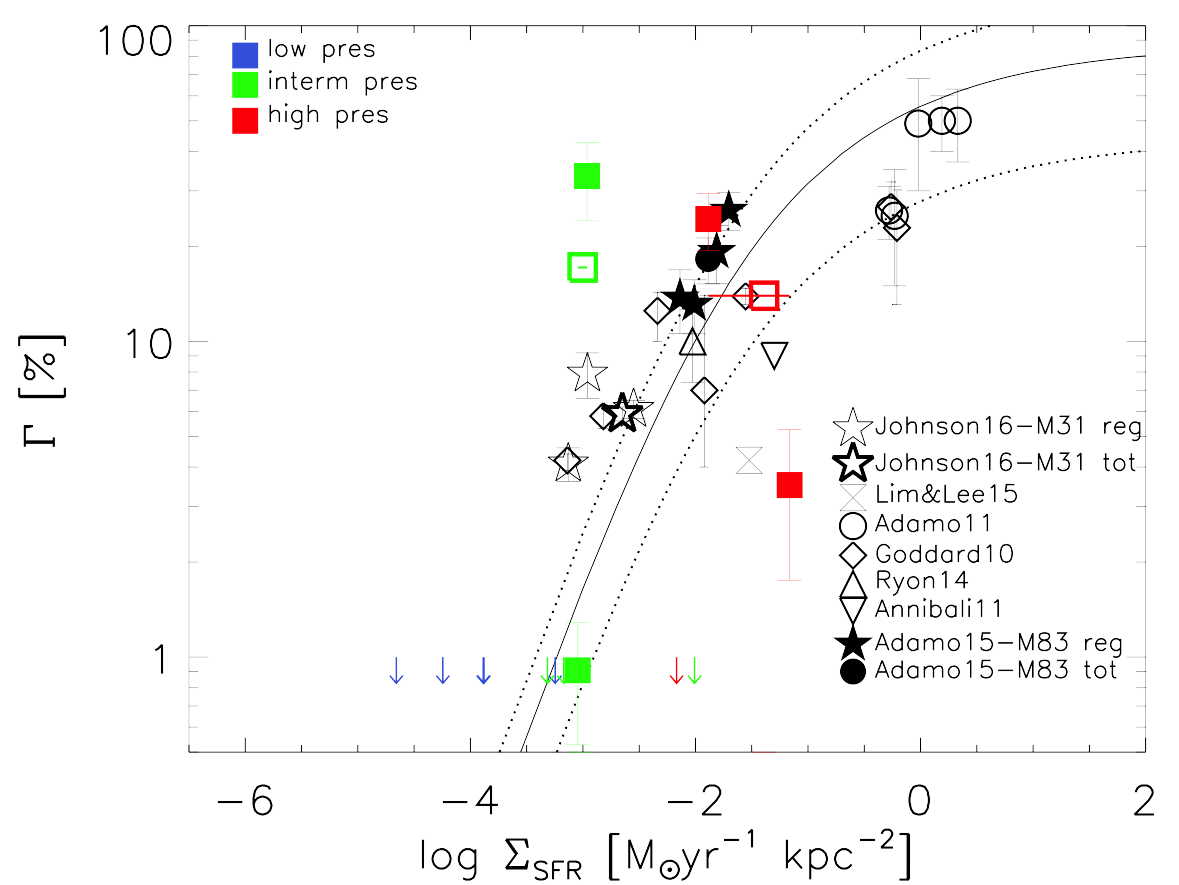}
\caption{
{\it Left:} $\Gamma$, the ratio of cluster formation rate to star formation rate, plotted on a logarithmic scale by pressure bin.
Pressure bin 1 is $\log$ pressure $<-12.4$,
bin 2 is $\log$ pressure between $-12.4$ and $-11.4$, and bin 3 is $\log$ pressure $>-11.4$.
Units of pressure g (s$^{2}$ cm)$^{-1}$.
The area included is the total area of pressure regions shown in Figure \ref{fig-pres} within the given bin range.
%DDO 63 does not contain any  clusters but is included in the labels as a reminder that its $\Gamma$ is 0.
%None of the galaxies has clusters in pressure bin 1.
{\it Right:} $\Gamma$ plotted on a logarithmic scale
as a function of the SFR per unit area in the pressure bin where $\Gamma$ was calculated.
The three pressure bins in each galaxy are plotted as filled squares color coded by pressure bin in the
upper left corner. The open squares are the averages of
the galaxies in that pressure bin. Pressure bins that have less than 2 clusters are shown as upper limits along the x axis.
Other samples are shown in black according to the legend on the bottom right of the plot.
The solid black line is the model of \citet{kruijssen12} for the formation of bound clusters and the dotted lines are the 1$\sigma$ uncertainties.}
\label{fig-gamma}
\end{figure}

\newpage
%fig22
\begin{figure}[t!]
\epsscale{1.0}
\vskip +1.5truein
%\plotone{fig22.eps}
\includegraphics[angle=0,width=1.0\textwidth]{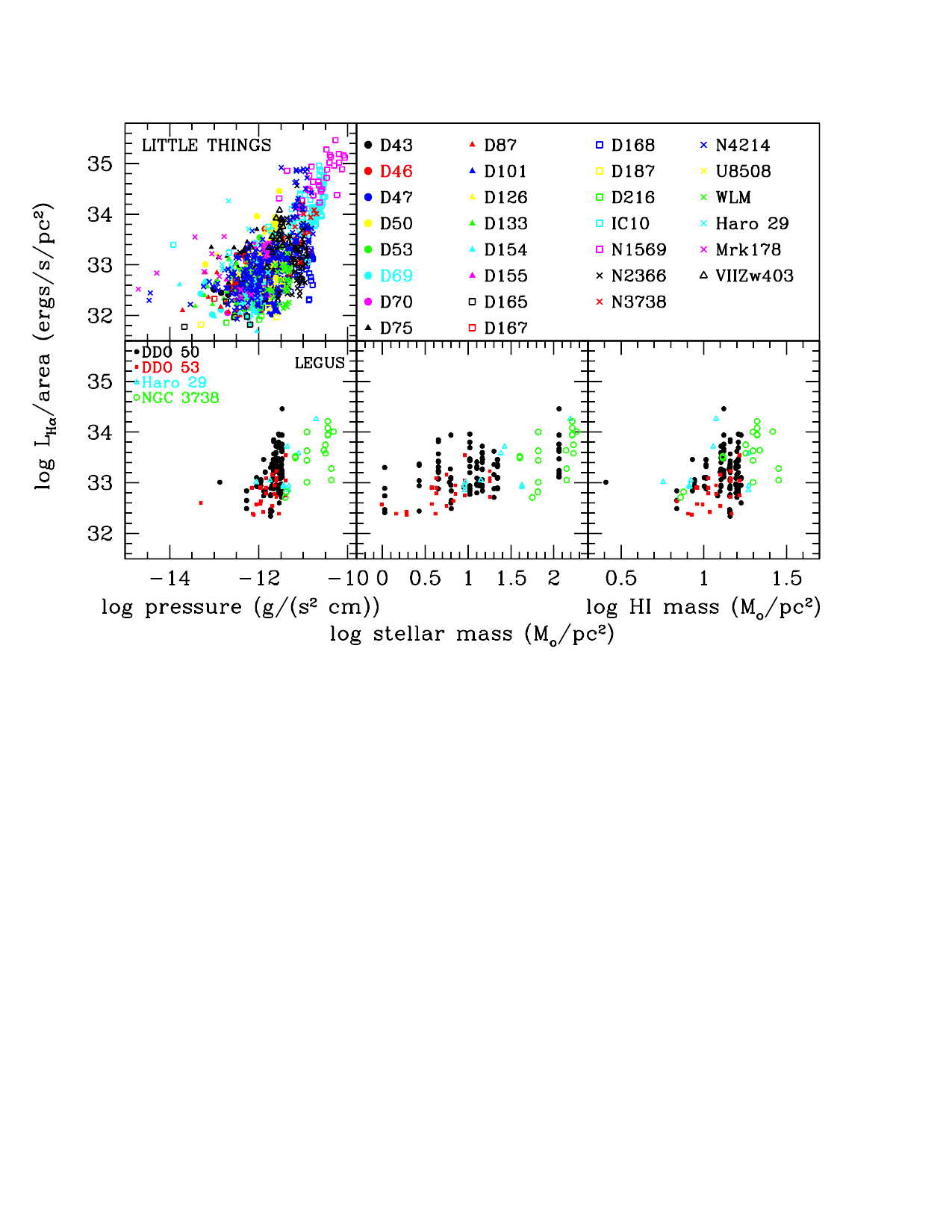}
\vskip -3.45truein
\caption{
\HII\ region \ha\ luminosity per unit area from \citet{hiilum} vs.\ surrounding galactic pressure. DDO 63 was not included in that sample.
{\it Top:} Twenty-nine of the LITTLE THINGS dIrr galaxies, including four from the sample concentrated on in this paper.
The pressure was measured in an annulus 200 pc wide beyond the \HII\ region.
{\it Bottom:} LEGUS/LITTLE THINGS dIrr sample only. The pressure was measured in regions defined by the pressure map as shown in
Figure \ref{fig-pres}.} %\HII\ regions and luminosities are from \citet{hiilum}
\label{fig-hiichar}
\end{figure}

\newpage
%fig23
\begin{figure}[t!]
\epsscale{0.95}
%\vskip 0.5truein
%\plotone{fig23.eps}
\includegraphics[angle=0,width=1.0\textwidth]{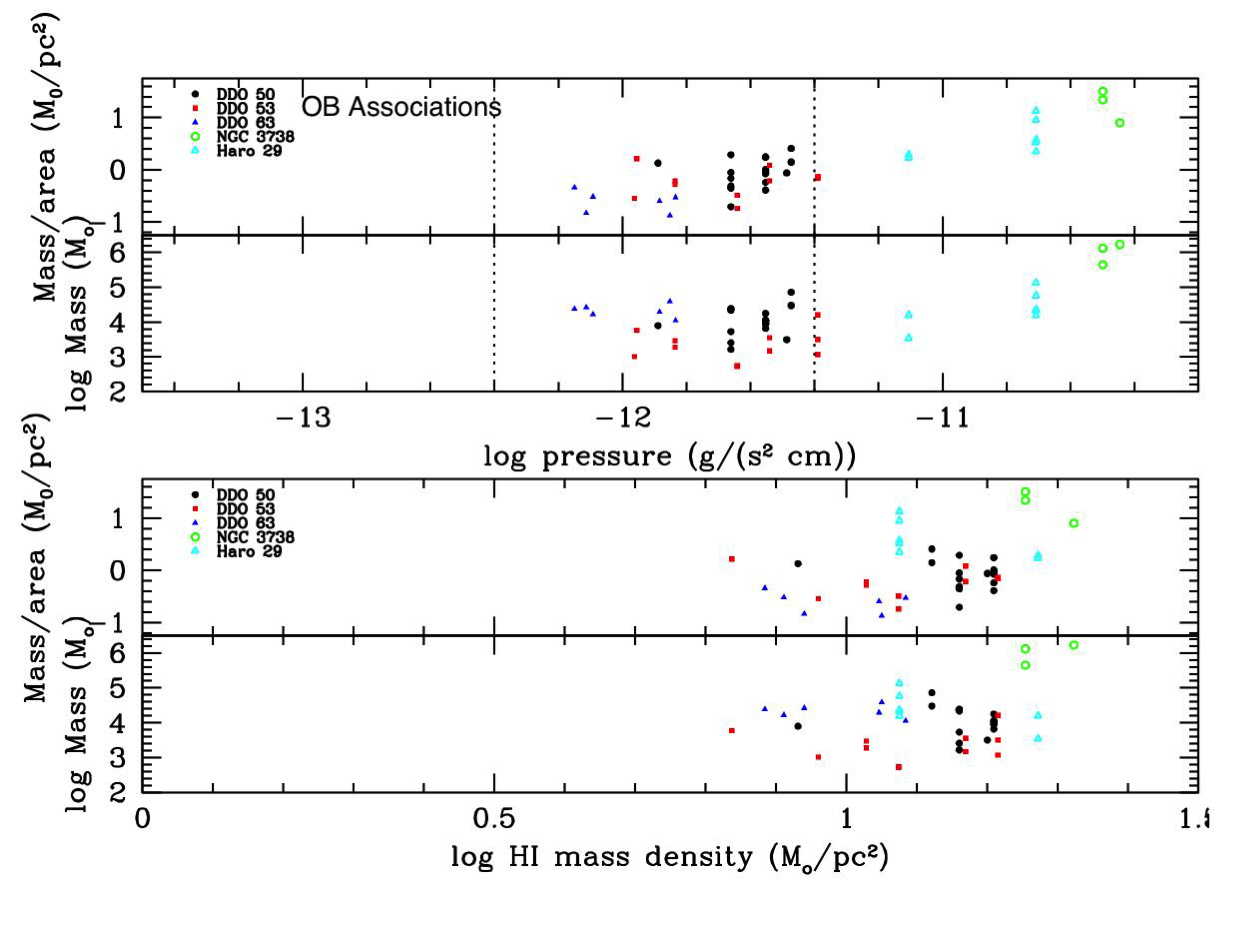}
\vskip -0.25truein
\caption{OB association characteristics of stellar mass and mass surface density vs.\
environmental characteristics of pressure and \HI\ surface density.
The vertical dotted lines in the top panel delineate the three pressure bins discussed in the text, and the
x-axis is the same as that of Figure \ref{fig-clenv} for  clusters.
We see that the OB associations in Haro 29 and NGC 3738 are more extreme in mass and mass density
than those in the other three dIrrs, and they are found at the highest pressure.}
\label{fig-obassocenv}
\end{figure}

\newpage
%fig24
\begin{figure}[t!]
\epsscale{0.95}
\vskip -0.25truein
%\plotone{fig24.eps}
\includegraphics[angle=0,width=1.0\textwidth]{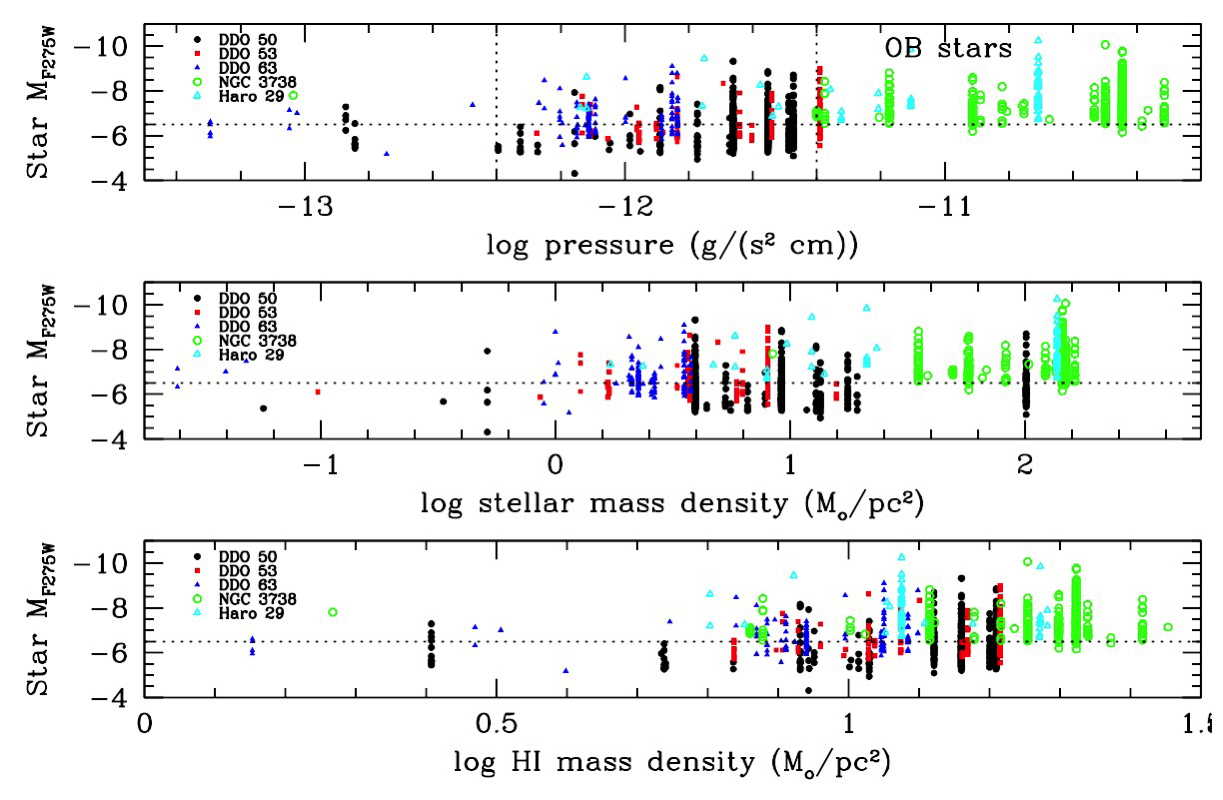}
\vskip -0.25truein
\caption{O star F275W absolute magnitude vs.\ galactic environment in which the stars are found.
Galactic environmental characteristics include pressure, stellar mass density, and \HI\ mass surface density.
The vertical dotted lines in the top panel delineate the three pressure bins discussed in the text, and the
x-axis is the same as that of Figure \ref{fig-clenv} for  clusters.
The variations in lower limits from galaxy to galaxy are likely due to distance effects
and incompleteness due to the higher backgrounds in the higher SFR galaxies.
The horizontal dashed line at an $M_{F275W}$ of $-6.5$ delineates an absolute magnitude cutoff that is common to all of the galaxies.
Note that we have ignored differential extinction within each galaxy and variations of average extinction between galaxies.
\label{fig-stenv}
}
\end{figure}

\newpage
%fig25
\begin{figure}[t!]
%\epsscale{0.75}
%\vskip 1.0truein
%\plotone{fig25.eps}
\includegraphics[angle=0,width=1.0\textwidth]{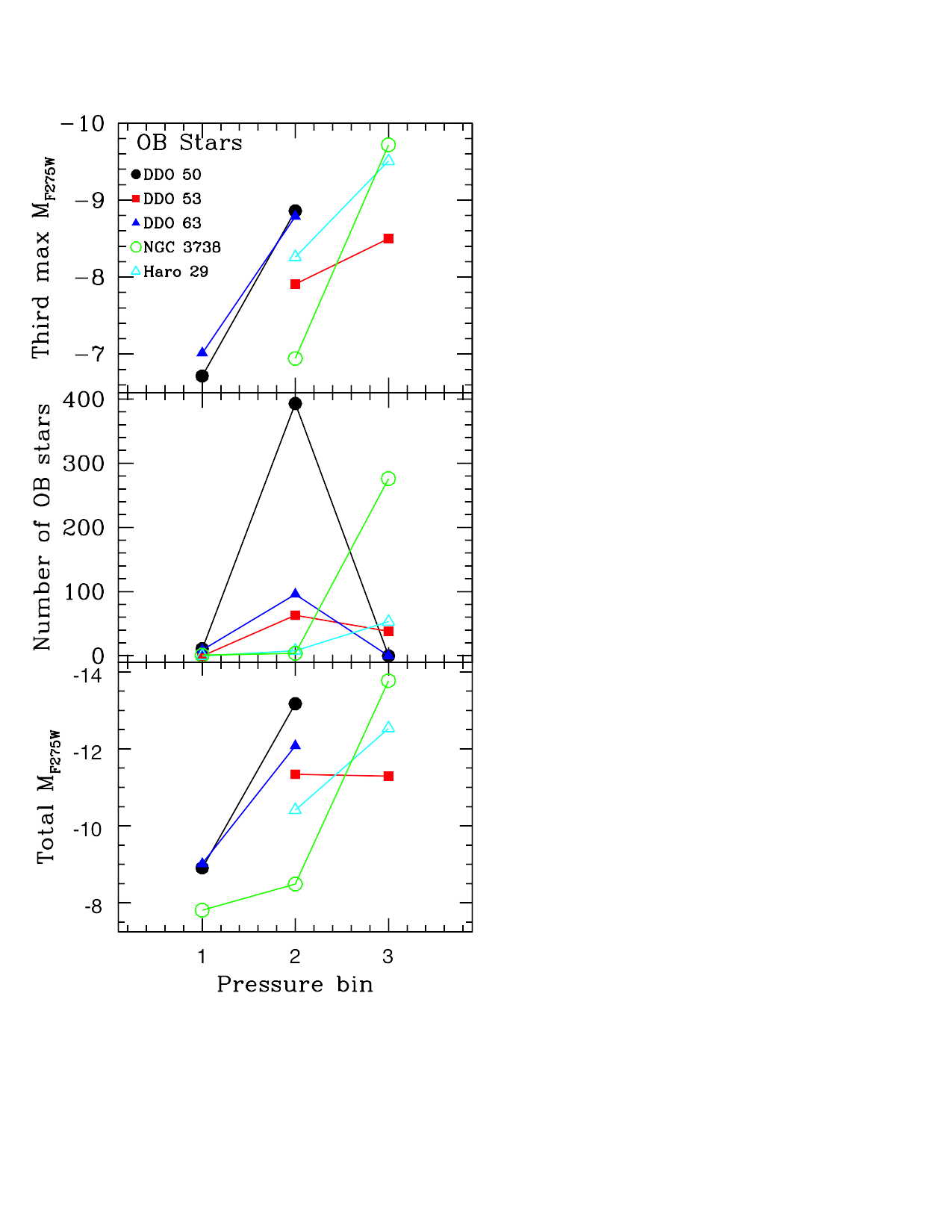}
\vskip -1.35truein
\caption{
O star characteristics by pressure bin:
third brightest absolute F275W magnitude, number of stars, and integrated stellar F275W absolute magnitude.
The pressures are combined in three bins; bin 1 is $\log$ pressure $<-12.4$,
bin 2 is $\log$ pressure between $-12.4$ and $-11.4$, and bin 3 is $\log$ pressure $>-11.4$.
Units of pressure g (s$^{2}$ cm)$^{-1}$.}
\label{fig-streg}
\end{figure}

\newpage
%fig26
\begin{figure}[t!]
%\epsscale{0.75}
%\vskip -0.25truein
%\plotone{fig26.eps}
\includegraphics[angle=0,width=1.0\textwidth]{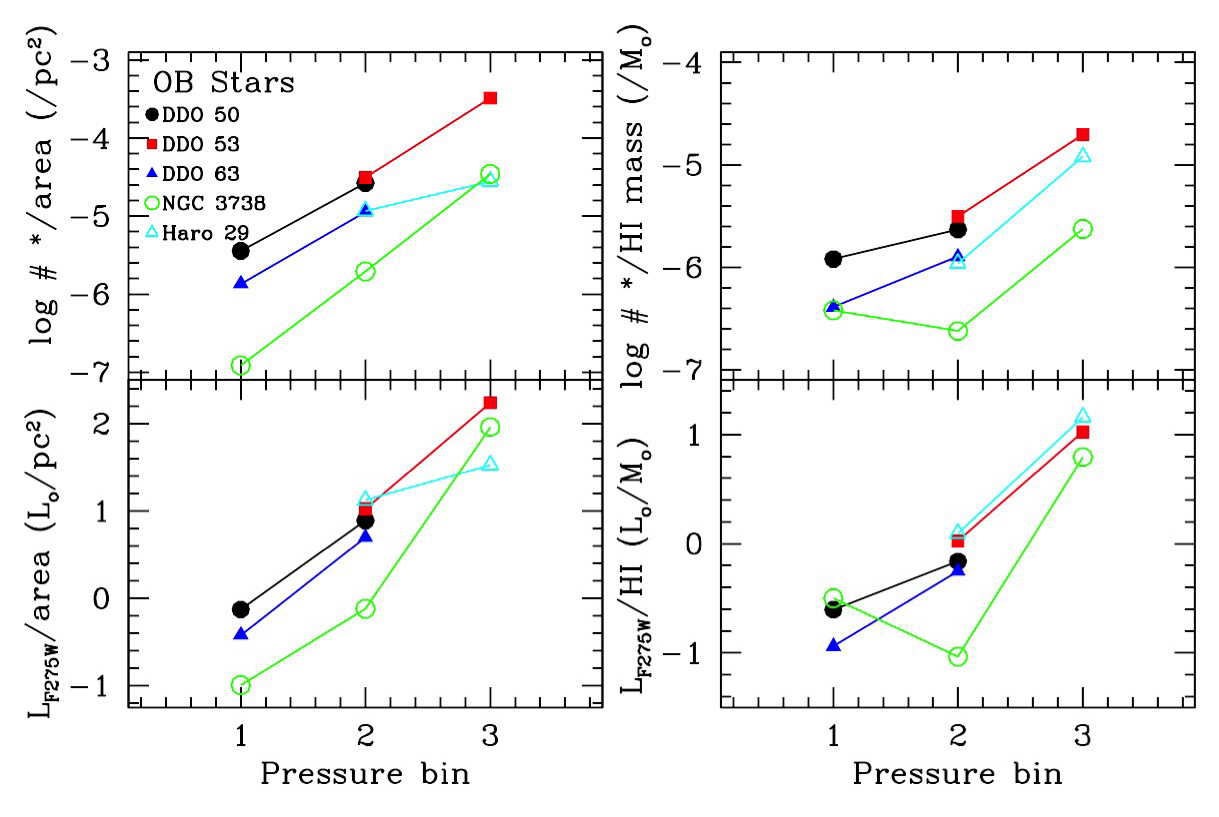}
%\vskip -0.25truein
\caption{
Number and total F275W luminosity of O stars per unit area and per \HI\ gas mass
vs.\ pressure in which the stars are found.
This is similar to Figure \ref{fig-clarea} but for O stars.
%; the top two panels are the same as in Figure \ref{fig-clarea}.
The pressures are combined in three bins; bin 1 is $\log$ pressure $<-12.4$,
bin 2 is $\log$ pressure between $-12.4$ and $-11.4$, and bin 3 is $\log$ pressure $>-11.4$.
Units of pressure g (s$^{2}$ cm)$^{-1}$, units of area are pc$^{2}$, the units of mass are M\solar,
and units of F275W luminosity are $L\solar$.
The area is the total area of pressure regions shown in Figure \ref{fig-pres} within the given bin;
similarly for the \HI\ mass.
Only NGC 3738 has O stars in all three pressure bins.}
\label{fig-starea}
\end{figure}

\newpage
%fig27
\begin{figure}[t!]
%\epsscale{0.6}
%\vskip +1.5truein
%\plotone{fig27.eps}
\includegraphics[angle=0,width=1.0\textwidth]{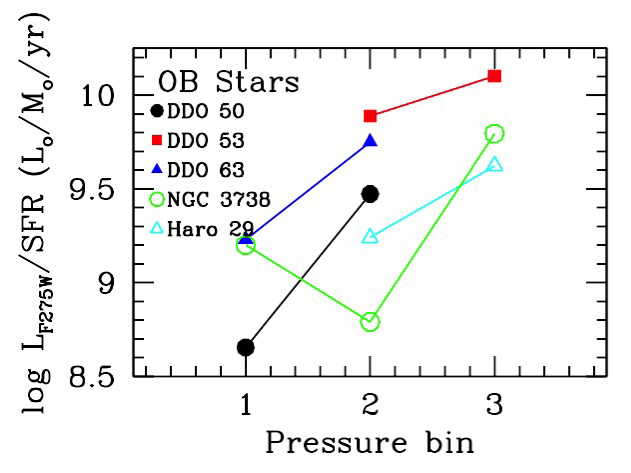}
%\vskip -0.25truein
\caption{
Total O star F275W luminosity divided by the SFR by pressure bin,
similar to Figure \ref{fig-clsfr} for  clusters.
The pressures within each galaxy are combined in three bins; bin 1 is $\log$ pressure $<-12.4$,
bin 2 is $\log$ pressure between $-12.4$ and $-11.4$, and bin 3 is $\log$ pressure $>-11.4$.
Units of pressure g (s$^{2}$ cm)$^{-1}$, units of SFR are M\solar\ yr$^{-1}$,
and units of F275W luminosity are $L\solar$.
Only NGC 3738 has O stars in all three pressure bins.}
\label{fig-stsfr}
\end{figure}

\newpage
%fig28
\begin{figure}[t!]
%\centering%
%\plottwo{fig28left.eps}{fig28right.eps}
%\begin{subfigure}{.5\textwidth}
%  \centering
%  %\includegraphics[width=1.0\linewidth]{fig28left.jpg}
%\end{subfigure}%
%\begin{subfigure}{.5\textwidth}
%  \centering
%  %\includegraphics[width=1.0\linewidth]{fig28right.jpg}
%\end{subfigure}
    %\subfloat[]{{\includegraphics[width=7.5cm]{fig28left.eps} }}%
    %\qquad
    %\subfloat[]{{\includegraphics[width=7.5cm]{fig28right.eps} }}%
    \includegraphics[width=7.5cm]{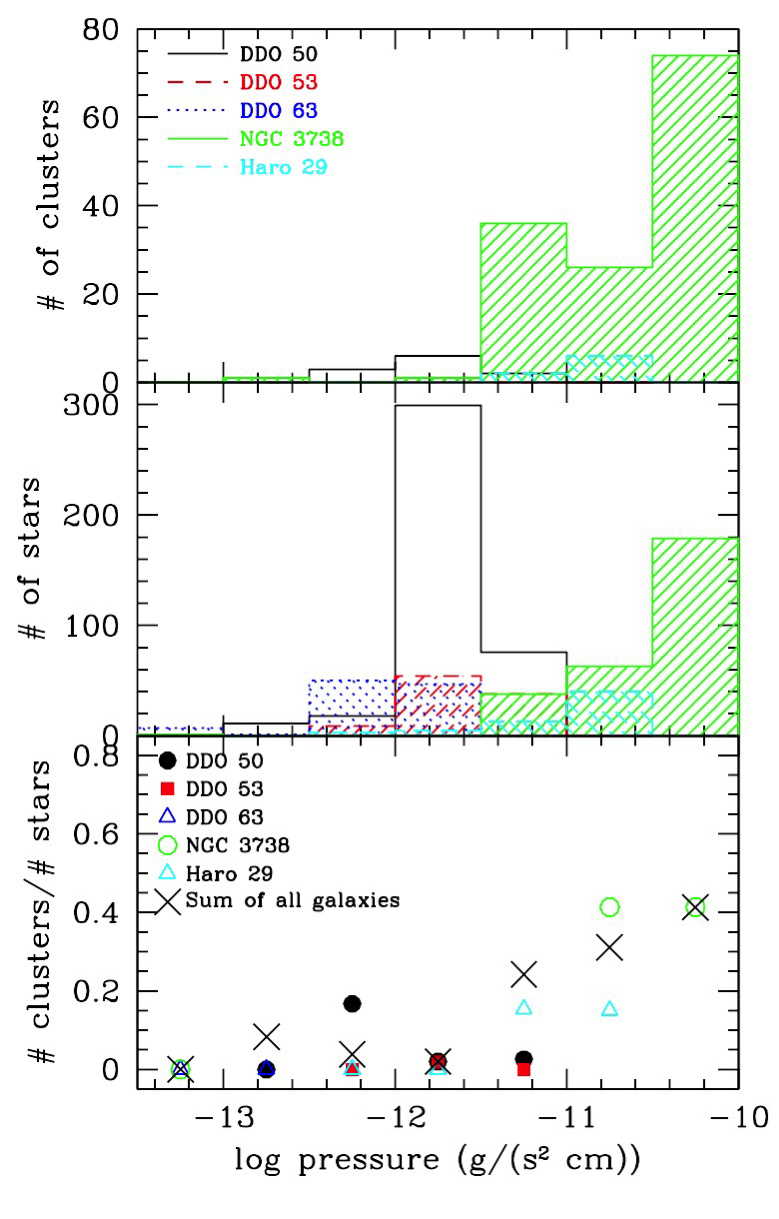}
    \includegraphics[width=7.5cm]{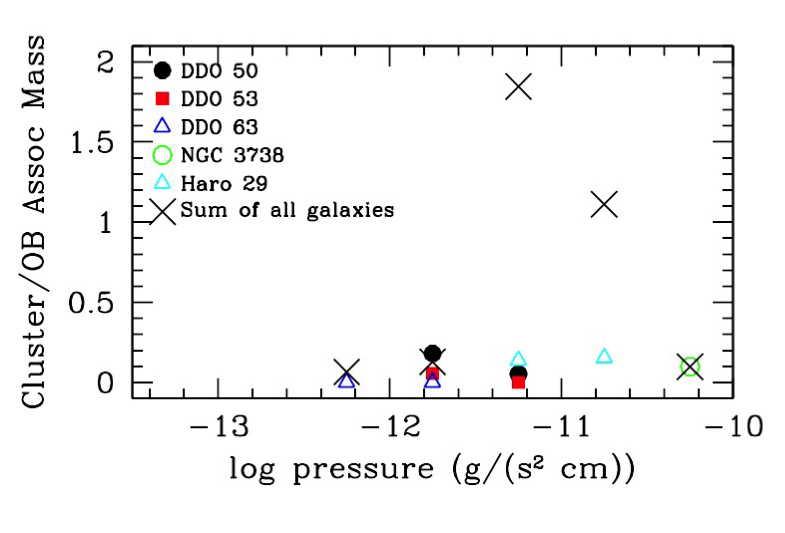}
\caption{
{\it Left}: Number of  clusters, number of O stars, and the ratio of clusters to O stars plotted against
the pressure binned by 0.5 in the logarithm.
The large black X in the bottom panel comes from adding all of the clusters and stars in each pressure bin in all 5 galaxies
and taking the ratio; it is not the average of the individual galaxy ratios.
{\it Right}: Total mass of clusters divided by the total mass of OB associations, versus pressure bin.
For a given pressure bin, a galaxy might have clusters but no OB associations or vice versa.
The large black X in the bottom panel comes from adding the mass of all of the clusters and OB
associations in each pressure bin in all 5 galaxies.}
\label{fig-rat}
\end{figure}

\end{document}